\documentclass[lineno=false]{jfm}

\usepackage{graphicx}
\usepackage{layouts}
\usepackage{pdflscape}
\usepackage{longtable}
\usepackage{newtxtext}
\usepackage{newtxmath}
\usepackage{natbib}
\usepackage{hyperref}
\hypersetup{
    colorlinks = true,
    urlcolor   = blue,
    citecolor  = black,
}

\newcommand{\RomanNumeralCaps}[1]
\linenumbers

\usepackage{placeins}
\usepackage[tableposition=below,width=\textwidth,justification=justified]{caption}

\newcommand{\ctr}[1]{\multicolumn{1}{c}{#1}}

\newcommand\cites[1]{\citeauthor{#1}'s\ (\citeyear{#1})}

\shorttitle{Scaling relations in sheared Rayleigh--B\'enard convection} 
\title{Scaling relations for heat and momentum transport in sheared Rayleigh--B\'enard convection}
\author
{
    Guru Sreevanshu Yerragolam\aff{1}
    \corresp{\email{g.s.yerragolam@utwente.nl}},
    Christopher J. Howland\aff{1},
    Richard J.A.M. Stevens\aff{1},
    Roberto Verzicco\aff{2,1,3},
    Olga Shishkina\aff{4} 
    \and 
    Detlef Lohse\aff{1,4},
    \corresp{\email{d.lohse@utwente.nl}}
}

\affiliation
{
    \aff{1}
    Physics of Fluids Group, Max Planck Center for Complex Fluid Dynamics, J. M. Burgers Center for Fluid Dynamics, Department of Science and Technology, University of Twente, P.O. Box 217, 7500 AE Enschede, The Netherlands
    \aff{2}
    Dipartimento di Ingegneria Industriale, University of Rome "Tor Vergata". Via del Politecnico 1, Roma 00133, Italy
    \aff{3}
    Gran Sasso Science Institute - Viale F. Crispi, 7 67100 L'Aquila, Italy.
    \aff{4}
    Max Planck Institute for Dynamics and Self-Organization, Am Fassberg 17, 37077 G\"ottingen, Germany
}

\begin{document}
\maketitle

\begin{abstract}

We provide scaling relations for the Nusselt number $Nu$ and the friction coefficient $C_{S}$ in sheared Rayleigh--B\'enard convection, i.e., in Rayleigh--B\'enard flow with Couette or Poiseuille type shear forcing, by extending the \citet{gro00,gro01,gro02,gro04} theory to sheared thermal convection. The control parameters for these systems are the Rayleigh number $Ra$, the Prandtl number $Pr$, and the Reynolds number $Re_S$ that characterises the strength of the imposed shear. By direct numerical simulations and theoretical considerations, we show that in turbulent Rayleigh--B\'enard convection, the friction coefficients associated with the applied shear and the shear generated by the large-scale convection rolls are both well described by \cites{pra32a} logarithmic friction law, suggesting some kind of universality between purely shear driven flows and thermal convection. These scaling relations hold well for $10^6 \leq Ra \leq 10^8$, $0.5 \leq Pr \leq 5.0$, and $0 \leq Re_S \leq 10^4$.
\end{abstract}

\section{Introduction}
\label{sec:intro}

\begin{figure}
    \centering
    \includegraphics[width=\textwidth]{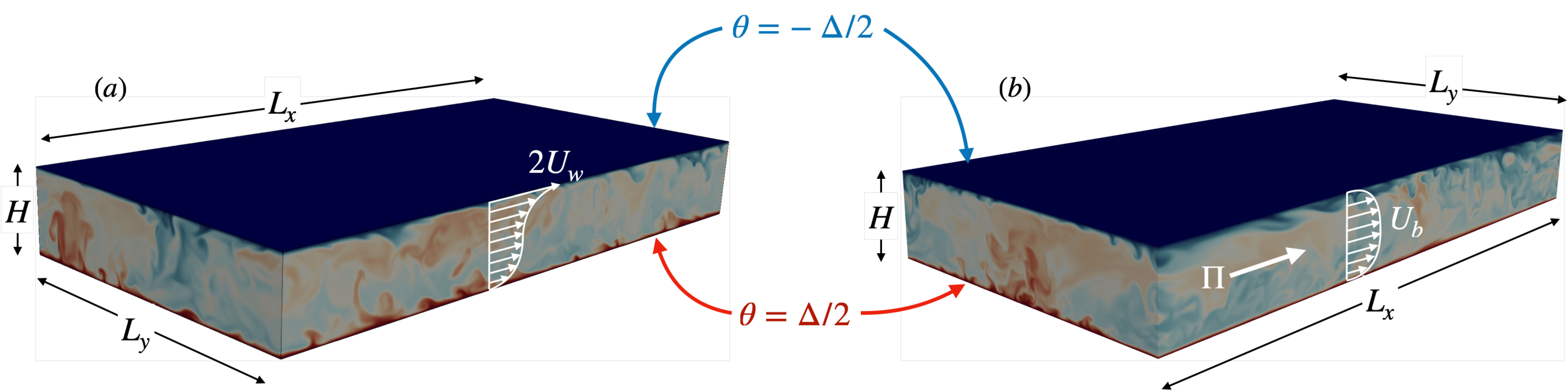}
    \caption{Schematic of the (a) Couette--Rayleigh--B\'enard (CRB) and (b) Poiseuille--Rayleigh--B\'enard (PRB) systems}
    \label{fig:schematic}
\end{figure}

The interplay between buoyancy and shear in mixed thermal convection can be studied by either adding Couette--type forcing to the Rayleigh--B\'enard (RB) system \citep{ahl09,loh10,chi12,xia13,shi21,ahl22,loh23} to obtain the Couette--RB (CRB) system \citep{dea65,ing66,hat86,dom88,sol90,she19,bla20,bla21b}, or by applying a Poiseuille--type forcing to obtain the Poiseuille--RB (PRB) system \citep{zon14,sca14,sca15,pir17}. A schematic of the two systems is shown in figure \ref{fig:schematic}. 
    
The CRB and PRB systems are described by the incompressible Navier--Stokes equations, the continuity equation, and the temperature transport equation, within the Boussinesq approximation. In Cartesian coordinates, they read
\begin{equation}
    {\partial_t u_i} + u_j {\partial_j u_i} = - \rho^{-1} {\partial_i p} + \nu {\partial^2_j u_i} + \beta g \delta_{i3}\theta + \Pi \delta_{i1}, \ \ \ \ \ {\partial_i u_i} = 0,
    \label{eqn:Navier}
\end{equation}
\begin{equation}
    {\partial_t \theta} + u_j {\partial_j \theta} = \kappa {\partial^2_j \theta},
    \label{eqn:temp}
\end{equation}
\noindent where $\boldsymbol{u} \equiv (u_x,u_y,u_z)$ is the velocity, $p$ the pressure, $\theta$ the reduced temperature, $\rho$ the density of the fluid, $g$ the acceleration due to gravity antiparallel to the $z$ direction, $\beta$ the isobaric thermal expansion coefficient, $\nu$ the kinematic viscosity, $\kappa$ the thermal diffusivity, and $H$ is the distance between the horizontal walls. At the top wall ($z=H$), the reduced temperature is set to $\theta = -\Delta/2$ while at the bottom wall ($z=0$), the reduced temperature is set to $\theta = \Delta/2$. For the CRB system $\Pi = 0$, the bottom wall is at rest and a velocity of $2U_w$ is imposed on the top wall. For the PRB system, no-slip conditions are enforced at the walls and a volume forcing $\Pi$ is applied in the streamwise direction such that it induces a bulk velocity of $U_b$ averaged over the domain volume and time (for the details of the implementation of the shear forcing in the numerical simulations, we refer the reader to \S\ref{sec:numerical}). The streamwise direction is oriented along $x$ and the spanwise direction along $y$. The aspect ratios of the system are defined by $\Gamma_x = L_x/H$ and $\Gamma_y = L_y/H$, with $L_x$, $L_y$ being the dimensions of the system in the $x$ and $y$ directions, respectively.

The control parameters for the systems are the Rayleigh number, the Prandtl number, and the Reynolds number associated with the shear forcing:
\begin{align}
    Ra &\equiv \frac{\beta g H^3 \Delta}{\nu \kappa}, &
    Pr &\equiv \frac{\nu}{\kappa}, &
    Re_S &\equiv \frac{U_S H}{\nu} .
    \label{eqn:input_params}
\end{align}
The characteristic velocity scale associated with the shear forcing $U_S$ is given by $U_S \equiv U_w$ for the CRB system and $U_S \equiv U_b$ for the PRB system. Although $U_b$ is formally a response parameter, in our numerical simulations the volume forcing term $\Pi$ is computed at each time-step to ensure a constant mass flow rate \citep{qua16} dictated by $U_b$, making it a control parameter in our case. The shear forcing for the CRB system is given by the wall Reynolds number $Re_w \equiv {U_w H}/{\nu}$ whereas the shear forcing for the PRB system is given by the bulk Reynolds number $Re_b \equiv {U_b H}/{\nu}$. Henceforth, we use $Re_S$ to indicate shear forcing in equations that are applicable to both CRB and PRB systems. 

Similarly, one can also define a Reynolds number associated with the thermal forcing in these systems. From the input parameters, we can construct a Reynolds number ${Re_F = U_F H /\nu \equiv \sqrt{Ra/Pr}}$, using the free fall velocity scale $U_F = \sqrt{g \beta \Delta H}$. However, the free fall scale is often not a reliable estimate of the flow velocities that develop in natural convection flows. A more appropriate approach is to define the Reynolds number associated with the large-scale convection (LSC) roll given by $Re_L \equiv U_L H / \nu $ with $U_L$ indicating the mean velocity of the ``wind of turbulence'' generated by the LSC roll. The parameter $Re_L$ is however, a response parameter whose variation with $Ra$, $Pr$, and $Re_S$ is not known \textit{a priori}. 

In the limiting case of zero shear forcing, we can distinguish the Reynolds number associated with the LSC roll in pure RB flow as $Re_R(Ra,Pr) \equiv Re_L(Ra,Pr,Re_S=0)$. The dynamics of the sheared RB systems are governed by a combined effect of both shear and thermal forcing. Therefore, we can also introduce the Reynolds number $Re_T$ which is constructed using the total velocity $U_T$ comprising a vector sum of $U_S$ and $U_L$. Naturally, the time averaged wall shear stress $\tau_T$ generated by the total velocity $U_T$ is also determined by the combined effect of $\tau_L$, which is the time averaged shear stress locally generated on the walls by the LSC roll, and $\tau_S$ which is the mean streamwise shear stress generated on the walls due to the applied shear forcing. These shear stresses can be expressed in dimensionless form using the friction coefficients associated with the total shear, the LSC and the streamwise shear respectively, as
\begin{align}
    C_T &\equiv \frac{2 \tau_T}{\rho U_T^2}, &
    C_L &\equiv \frac{2 \tau_L}{\rho U_L^2}, &
    C_S &\equiv \frac{2 \tau_S}{\rho U_S^2}. \label{eqn:friction_def}
\end{align}
Once again, in the limiting case of zero shear forcing, we can distinguish the friction coefficient associated with the LSC roll in pure RB flow as $C_R(Ra,Pr) \equiv C_L(Ra,Pr,Re_S=0)$. The non-dimensional heat flux from the hot bottom wall to the cold top wall is given by the Nusselt number:
\begin{equation}
    Nu \equiv \frac{\left< u_z \theta - \kappa \partial_z \theta \right>_{A,t}} {\kappa \Delta H^{-1}}, \label{eqn:Nusselt_def}
\end{equation}
with $\left< \right>_{A,t}$ indicating the averaging in time and over any horizontal plane spanned by $x$ and $y$. A summary of all response and control parameters discussed above is given in table \ref{tab:parameters} for reference.
In this study, we are primarily interested in understanding the dependence of the response parameters on the control parameters, and the physics underlying the connections between the response parameters.

\begin{table}
    \centering
    \begin{tabular}{ c c c p{65mm} }
        Parameter & Definition & Type & Short description \\
        \hline
        
        $Ra$ & $\bigl(\beta g H^3 \Delta\bigr)/(\nu \kappa)$ & Control & Rayleigh number \\
        $Pr$ & $\nu/\kappa$ & Control & Prandtl number \\
        
        $Re_F$ & $U_F H/\nu \equiv \sqrt{Ra/Pr}$ & Control & Reynolds number associated with free-fall velocity \\
        
        $Re_w$ & $U_w H/\nu$ & Control & Wall Reynolds number for CRB system \\
        $Re_b$ & $U_b H/\nu$ & Control & Bulk Reynolds number for PRB system \\
        
        $Re_S$ & $U_S H/\nu$ & Control & Reynolds number associated with shear forcing \\
        $Re_L$ & $U_L H/\nu$ & Response & Reynolds number associated with LSC rolls for sheared RB\\
        $Re_R$ & $U_R H/\nu \equiv Re_L(Re_S=0)$ & Response & Reynolds number associated with LSC rolls for pure RB (``wind Reynolds number'') \\
        $Re_T$ & $U_T H/\nu$ & Response & Reynolds number computed using the total velocity\\

        $C_S$ & $2 \tau_S/\bigl(\rho U_S^2\bigr)$ & Response & Friction coefficient associated with shear forcing \\
        $C_L$ & $2 \tau_L/\bigl(\rho U_L^2\bigr)$ & Response & Friction coefficient associated with LSC rolls for sheared RB\\
        $C_R$ & $2 \tau_R/\bigl(\rho U_R^2\bigr) \equiv C_L(Re_S=0)$ & Response & Friction coefficient associated with LSC rolls for pure RB\\
        $C_T$ & $2 \tau_T/\bigl(\rho U_T^2\bigr)$ & Response & Friction coefficient computed using the total velocity\\

        $Nu$ & $\left< u_z \theta - \kappa \partial_z \theta \right>_{A,t} / \bigl(\kappa \Delta H^{-1}\bigr)$ & Response & Nusselt number \\

        
    \end{tabular}
    \caption{A summary of all control and response parameters discussed in \S\ref{sec:intro}.}
    \label{tab:parameters}
\end{table}

Using the exact relations for the global kinetic and thermal dissipation rates, \citet{gro00,gro01,gro02,gro04} offered a unifying theory for RB convection (hereafter referred to as the GL-theory). For cylinders of unit aspect ratio, \citet{ste13} have demonstrated that fitting the GL-theory at four data points from \citet{fun05} at $Ra = 2.96 \times 10^7$ and $Ra = 1.92 \times 10^{10}$ with $Pr = 4.38$, from \citet{xia02} at $Ra = 2.24 \times 10^8$ with $Pr = 554$ and from \citet{ker00} at $Ra = 10^7$ with $Pr = 0.07$ using four free parameters can predict $Nu$ within $4\%$ of experimental and numerical results in most of the parameter space given by $10^{4} \leq Ra \leq 10^{14}$ and $10^{-3} \leq Pr \leq 10^{2}$ with only two small ranges that exhibit a greater $10\%$ disagreement. In the same paper \citet{ste13}, a fit of similar quality is achieved also for an aspect ratio Gamma = 1/2, with slightly different prefactors. This work has further been extended by \citet{ahl22} to include the effects of aspect ratio between $1/32$ and $32$. The GL-theory has also been extended to the ultimate regime \citep{gro11}, where the heat transport is considerably enhanced, as the laminar-type boundary layers become turbulent due to a non-normal--nonlinear instability; for a detailed discussion see \citet{roc20} and \citet{loh23}. 

Presently, the GL-theory has been applied to RB convection without imposed shear. The objective of this work is to extend the theoretical approach to sheared thermal convection. \citet{sca14,sca15,pir17,bla20,bla21b} have made progress in understanding the variation of heat transfer in RB convection with imposed shear. \citet{sca14,sca15} proposed a model based on the concept of eddy viscosity and eddy diffusivity to explain the counter-intuitive initial decrease and subsequent increase in $Nu$ with increasing $Re_b$ for the PRB system. \citet{bla20} observed a similar effect in the CRB system and attributed it to the initial destruction of the large-scale flow organisation and the subsequent formation of large meandering flow structures \citep{hut07}. They divided the flow into a buoyancy-dominated, a transitional, and a shear-dominated regime, based on the Monin--Obukhov length scale. \citet{bla21b} further investigated the effect of $Pr$ on the variation of $Nu$ with $Re_w$ and concluded that the non-monotonic behaviour of $Nu(Re_w)$ is a consequence of flow layering, plume sweeping, and bulk heat entrapment. Building on these findings, in this paper we propose a more general formulation applicable to sheared thermal convection, in the spirit of the GL-theory.

The objective of this paper is to extend the GL-theory to CRB and PRB systems by providing scaling relations for $Nu(Ra,Pr,Re_S)$ and $C_S(Ra,Pr,Re_S)$. Furthermore, we will show similarities in the dependence of $C_S$ on $Re_S$ in Couette and Poiseuille flows and $C_L$ on $Re_L$ in RB convection, suggesting some sort of universality in the behaviour of the friction coefficient in shear-driven flows and thermal convection. In \S\ref{sec:theory}, we build the theoretical framework, using rigorous relations for globally averaged kinetic and thermal dissipation rates. In \S\ref{sec:numerical}, we validate the theoretical scaling relations against direct numerical simulations (DNS). Finally, the conclusions are presented in \S\ref{sec:conclusion}.

\section{Extending the Grossmann--Lohse theory to CRB and PRB.}
\label{sec:theory}

\subsection{Kinetic and thermal dissipation rates}
\label{subsec:diss}

To extend the GL-theory to sheared thermal convection, we formulate exact global relations for the kinetic $(\epsilon_u)$ and thermal $(\epsilon_\theta)$ dissipation rates in the CRB and PRB systems. These arise from the time- and volume-averaged equations for the kinetic energy and temperature variance, respectively. The relation for the mean thermal dissipation rate is the same as in the classical RB convection,
\begin{equation}
    \epsilon_\theta = \left< \kappa {(\partial_j \theta)}^2 \right>_{V,t} = \kappa\frac{\Delta^2}{H^2}Nu,
    \label{eqn:thermal_diss_srb}
\end{equation}
see e.g. \citet{shr90,sig94}.
The relation for the mean kinetic dissipation rate reads:
\begin{equation}
    \epsilon_u = \left< \nu {(\partial_j u_i)}^2 \right>_{V,t} = \frac{\nu^3}{H^4}\Bigl( \underbrace{(Nu-1) Ra Pr^{-2}}_{\text{Buoyancy term}} + \underbrace{C_S Re_S^3}_{\text{Shear term}}\Bigr),
    \label{eqn:kinetic_diss_srb}
\end{equation} 
\noindent with $\left< ... \right>_{V,t}$ indicating the average over time and volume. Note that the expression for thermal dissipation (\ref{eqn:thermal_diss_srb}) is the same as that in classical RB convection but the expression for kinetic dissipation (\ref{eqn:kinetic_diss_srb}) includes contributions from both buoyancy and shear forcing.

\subsection{Kinetic energy, large scale circulation, and boundary layer thickness}
\label{subsec:lsc}

One of the central ideas of the GL-theory is the presence of persistent large-scale circulation (LSC) rolls that churn the bulk of the system and generate boundary layers at the walls. As a result, the mean kinetic energy of the RB system is expected to scale as $\sim U_R^2$, where $U_R$ is the velocity scale of the LSC (in the absence of any shear forcing). In pure RB flow, this mean kinetic energy is solely generated by the buoyancy forcing. However, in the case of sheared RB flow, where the LSC has a velocity scale $U_L$, the mean kinetic energy consists of contributions from both the LSC and the imposed shear flow.
Therefore, we add the kinetic energy of the mean flow $U_S^2$ and the associated turbulent kinetic energy (TKE), which scales as the square of the friction velocity $u_\tau=\sqrt{\tau_S/\rho}$,
to write
\begin{equation}
    U_T^2 \approx U_L^2 + U_S^2 + 2\gamma u_\tau^2,
    \label{eqn:U_T}
\end{equation}
\noindent with $\gamma$ being a prefactor. This can also be written in terms of the corresponding Reynolds numbers as
\begin{equation}
    Re_T^2 \approx Re_L^2 + Re_S^2 + \gamma C_S Re_S^2.
    \label{eqn:Re_T}
\end{equation}
For a laminar Prandtl--Blasius \citep{pra04,bla08} type boundary layer, there is no TKE, so the contribution $C_S Re_S^2$ vanishes and we can approximate \eqref{eqn:Re_T} as 
\begin{equation}
    Re_T^2 \approx Re_L^2 + Re_S^2.
    \label{eqn:Re_T_approx}
\end{equation}
An interpretation of the above equation (\ref{eqn:Re_T_approx}) is that the velocity associated with the LSC is preferentially oriented along the direction orthogonal to the shear forcing, consistent with the flow structures observed by \citet{pir17,bla20}. Note that the validity of equation (\ref{eqn:Re_T_approx}) is limited to sufficiently low values of $Re_S$ wherein the contribution to spanwise shear stresses from shear forcing is negligible in comparison to the contribution from the LSC rolls. At high shear forcing, the spanwise shear stresses generated by velocity fluctuations arising purely from shear forcing may no longer be negligible, in which case equation (\ref{eqn:Re_T_approx}) no longer holds.

In the buoyancy-dominated regime, relation (\ref{eqn:Re_T_approx}) can be better understood by considering many LSC rolls each orientated at an angle $\alpha$ with the streamwise direction and studying the probability distribution of $\alpha$ given by $\phi(\alpha)$. Here, we mean LSC rolls in a broad sense, without addressing the exact details of the flow organisation at this stage. Empirical observations regarding flow organisation are reported in section \S\ref{subsec:lsc_results}.
Since the total velocity $U_T$ arises from a vector addition of the shear velocity and the LSC velocities, we can express it in terms of $\phi(\alpha)$ as
\begin{equation}
    U_T^2 = \int_{0}^{2\pi} { \phi(\alpha) \left(U_S^2 + U_L^2 + 2 U_S U_L \cos(\alpha) \right) d\alpha}.
    \label{eqn:total_velocity_pdf}
\end{equation}
Due to the symmetry of the system and the periodic boundary conditions in the horizontal directions, it is reasonable to consider that there are an equal number of clockwise and counter-clockwise LSC rolls within the sheared RB system. When averaged over the entire volume of the system, we postulate that the probability distribution $\phi(\alpha)$ is symmetric about the spanwise direction, i.e., about $\alpha = \pm \pi/2$. Applying this symmetry condition to (\ref{eqn:total_velocity_pdf}) gives us the relation (\ref{eqn:Re_T_approx}). The velocity scale $U_T$ associated with the kinetic energy of the system can thus be considered as a vector sum of perpendicular velocity contributions from the shear forcing in the streamwise direction and the LSC roll in the spanwise direction. Following this approach, we can also decompose the total shear stress $\tau_T$ generated by $U_T$ into the streamwise component $\tau_s = \tau_T U_S/U_T$, generated by $U_S$, and the spanwise component $\tau_L = \tau_T U_L/U_T$, generated by $U_L$. This assumption gives us three equivalent definitions of the kinetic boundary layer thickness $\lambda_u$, namely
\begin{equation}
    \lambda_u \equiv \frac{2H}{C_L Re_L} = \frac{2H}{C_S Re_S} = \frac{2H}{C_T Re_T}.
    \label{eqn:kinetic_bl}
\end{equation}
Here we used the slope criterion from \citet{shi10} for the definition of the kinetic boundary layer thickness. Similarly, we define the thermal boundary layer thickness $\lambda_\theta$ also with the slope criterion as 
\begin{equation}
    \lambda_\theta \equiv \frac{H}{2Nu}.
    \label{eqn:thermal_bl}
\end{equation}

\subsection{Estimating bulk and boundary layer contributions}
\label{subsec:estimate}

Another key idea of the GL-theory is the splitting of $\epsilon_u$ and $\epsilon_\theta$ into their bulk and boundary layer contributions as follows,
\begin{align}
    \epsilon_u &= \epsilon_{u,BL} + \epsilon_{u,bulk}, & \epsilon_\theta &= \epsilon_{\theta,BL} + \epsilon_{\theta,bulk},
    \label{eqn:dissipation_bl_bulk}
\end{align}
with $\epsilon_{u,BL}$ being the boundary layer contribution to the kinetic dissipation, $\epsilon_{u,bulk}$ being the bulk contribution to the kinetic dissipation, $\epsilon_{\theta,BL}$ being the boundary layer contribution to the thermal dissipation, and $\epsilon_{\theta,bulk}$ being the bulk contribution to the thermal dissipation. Therefore, we focus on estimating $\epsilon_{u,BL}$, $\epsilon_{u,bulk}$, $\epsilon_{\theta,BL}$, and $\epsilon_{\theta,bulk}$ for sheared RB systems.

First, we estimate $\epsilon_{u,BL}$ using (\ref{eqn:kinetic_bl}) as 
\begin{equation}
    \epsilon_{u,BL} \sim \nu \frac{U_T^2}{\lambda_u^2} \frac{\lambda_u}{H} \sim \frac{\nu^3}{H^4} C_T Re_T^3 \equiv \frac{\nu^3}{H^4} \bigl(\underbrace{C_L Re_L^3}_{\text{LSC term}} + \underbrace{C_S Re_S^3}_{\text{Shear term}}\bigr),
    \label{eqn:kdiss_bl}
\end{equation}
which is, in turn, a sum of contributions from the LSC and the applied shear forcing.
Note that the shear contribution here exactly matches that in the global relation \eqref{eqn:kinetic_diss_srb}.
Next, we estimate $\epsilon_{u,bulk}$. Here, it is important to note that the bulk dissipation rate is dominated by the contribution from LSC rolls, while the contribution from applied shear forcing is much smaller. Therefore, we only focus on estimating the contribution to $\epsilon_{u,bulk}$ from the LSC rolls by assume that the velocity scale $U_L$ associated with the LSC rolls is responsible for stirring the bulk with a kinetic energy that scales with $U_L^2$. However, LSC rolls are swept at the boundary layer height due to the applied shear forcing and the timescale of the stirring process is governed not by the velocity $U_L$ associated with the LSC but by $U_T$ which is the total velocity scale. As we shall find later in \S\ref{subsec:nu_retau_results}, this is a key assumption that explains the trend of $Nu$ with increasing $Re_S$ in the buoyancy-dominated regime. Following these assumptions, we write
\begin{equation}
    \epsilon_{u,bulk} \sim U_L^2 \frac{U_T}{H} \equiv  Re_L^2 Re_T.
    \label{eqn:kdiss_bulk}
\end{equation}
At high-shear forcing, the contribution of the boundary layer to the dissipation rate arising from shear forcing dominates the bulk contribution from LSC rolls. In the limiting case of passive transport in Couette/Poiseuille flow, one can rigorously derive that the total kinetic dissipation rate $\epsilon_u = (\nu^3/H^4) C_S Re_S^3$. In this sense, the kinetic dissipation rate of sheared RB at high shear forcing will always be dominated by the boundary layer contribution and estimating $\epsilon_{u,bulk}$ in the shear-dominated regimes is redundant.

The analogous estimate for $\epsilon_{\theta,BL}$ 
\begin{equation}
    \epsilon_{\theta,BL} \sim \kappa \frac{\Delta^2}{\lambda_\theta^2} \frac{\lambda_\theta}{H} \sim \kappa\frac{\Delta^2}{H^2} Nu,
    \label{eqn:tdiss_bl}
\end{equation}
\noindent is identical to the exact relation (\ref{eqn:thermal_diss_srb}), on the one hand showing consistency of the approach, but on the other hand not giving new information. Therefore, following the GL-theory, we match the magnitude of the advective and diffusive terms of (\ref{eqn:temp}) at the thermal boundary layer height to obtain 
\begin{equation}
    u_y \partial_y \sim \kappa \partial_{zz}.
    \label{eqn:tbl_equation}
\end{equation}
As in the GL-theory, for regimes where the thermal boundary layer is thicker than the kinetic boundary layer ($\lambda_\theta > \lambda_u$, associated with low $Pr$), we estimate $u_y \sim U_L$, $\partial_y \sim 1/H$, and $\partial_{z z} \sim \lambda_\theta^{-2}$. Using these estimates in (\ref{eqn:tbl_equation}) with $\lambda_u$ from (\ref{eqn:kinetic_bl}) and $\lambda_\theta$ from (\ref{eqn:thermal_bl}), we obtain
\begin{equation}
    Nu \sim Pr^{1/2} Re_L^{1/2} \equiv \frac{Pr^{1/2}}{C_L Re_L^{1/2}} C_T Re_T.
    \label{eqn:tbl_equation_l}
\end{equation}
For high $Pr$ regimes, where $\lambda_\theta < \lambda_u$, we estimate $u_y \sim U_L \lambda_\theta / \lambda_u$ due to the fact that the relevant velocity scale at the thermal boundary height is smaller than the velocity $U_L$ by a factor $\lambda_\theta / \lambda_u$, exactly as in the GL-theory. Once again using the estimates $\partial_y \sim 1/H$, and $\partial_{z z} \sim \lambda_\theta^{-2}$ in (\ref{eqn:tbl_equation}) with $\lambda_u$ from (\ref{eqn:kinetic_bl}) and $\lambda_\theta$ from (\ref{eqn:thermal_bl}), we get
\begin{equation}
    Nu \sim Pr^{1/3} Re_L^{1/3} \left(C_T Re_T\right)^{1/3} \equiv \frac{Pr^{1/3}}{C_L^{2/3} Re_L^{1/3}} C_T Re_T.
    \label{eqn:tbl_equation_u}
\end{equation}
Also the bulk contribution to the thermal dissipation is estimated in an identical manner to the corresponding equation in the GL-theory, only with minor changes to reflect the new dependence on $C_L$, $C_T$, and $Re_T$. For $\lambda_\theta > \lambda_u$,
\begin{equation}
    \epsilon_{\theta,bulk} \sim \Delta^2\frac{U_L}{H} \sim \kappa\frac{\Delta^2}{H^2} Pr Re_L \equiv \kappa\frac{\Delta^2}{H^2} \frac{Pr}{C_L} \left(C_T Re_T\right),
    \label{eqn:tdiss_bulk_l}
\end{equation}
and for $\lambda_\theta < \lambda_u$, where the relevant velocity scale is $U_L \lambda_\theta/\lambda_u$, we get 
\begin{equation}
    \epsilon_{\theta,bulk} \sim \Delta^2\frac{U_L}{H} \frac{\lambda_\theta}{\lambda_u} \sim \kappa\frac{\Delta^2}{H^2} Pr C_L Re_L^2 Nu^{-1} \equiv \kappa\frac{\Delta^2}{H^2} \frac{Pr}{C_L} \frac{\left(C_T Re_T\right)^2}{Nu}.
    \label{eqn:tdiss_bulk_u}
\end{equation}
We expect equations (\ref{eqn:tbl_equation_l}) - (\ref{eqn:tdiss_bulk_u}) to be valid in both buoyancy-dominated and shear-dominated regimes. This is accommodated by the fact that the dependencies $C_T(Re_T)$ and $C_L(Re_L)$ behave differently in the buoyancy-dominated and shear-dominated regimes. Compatibility of these relations in the limiting regimes is explained in the subsequent subsection \S\ref{subsec:regimes}.

\subsection{Limiting regimes}
\label{subsec:regimes}

It is important to note that there exist no pure power laws for $Nu$, $Re_L$, and $C_S$ as functions of $Ra$, $Pr$, and $Re_S$. Nonetheless, it is useful to study the pure scaling power laws that arise from limiting regimes, in the interest of understanding the physics of the system. Based on the dominance of boundary layer and bulk contributions to the kinetic and thermal dissipation rates, the GL-theory provides four regimes $I$, $II$, $III$, and $IV$ for the pure RB system. Furthermore, each regime can be divided into two subregimes based on whether the thermal boundary layer is nested into the kinetic boundary layer or vice versa. We find that this classification of regimes is also applicable to buoyancy-dominated sheared RB. The phase space of $Ra$, $Pr$, and $Re_S$ is divided by four different transitions - (i) transition from boundary layer-dominated regimes to bulk-dominated regimes, (ii) transition from buoyancy-dominated regime to shear-dominated regime, (iii) transition between regimes where the thermal boundary layer is nested inside the kinetic boundary layer or vice versa, and (iv) transition from a laminar Prandtl--Blasius type boundary layer to a turbulent Prandtl--von K\'arm\'an type boundary layer. 

It should also be noted that some of these limiting regimes may not exist for sheared RB in the shear-dominated state. For example, at high shear forcing, the shear term in $\epsilon_{u,BL}$ always dominates $\epsilon_u$ because shear forcing primarily increases the boundary layer contribution of the kinetic dissipation $\left(\nu^3/H^4\right)C_S Re_S^3$. Since the friction coefficient $C_S$ associated with the applied shear can become independent of $Re_S$ only asymptotically at infinite $Re_S$, shear-dominated systems with $\epsilon_u \sim \epsilon_{u,bulk}$ can not exist.

We now proceed to first analyse the shear-dominated regime with $Re_S/Re_L \gg 1$. This can be realised in two ways -- namely either $Re_L$ is small or $Re_L$ is not necessarily small but $Re_S \gg Re_L$. When $Re_L$ is small, we assume the existence of a laminar Prandtl--Blasius type boundary layer with 
\begin{equation}
    C_L \sim Re_L^{-1/2},
    \label{eqn:cfl_Prandtl_Blasius}
\end{equation}
\noindent which, along with (\ref{eqn:kinetic_bl}) and (\ref{eqn:cfl_Prandtl_Blasius}) gives
\begin{equation}
    C_S \sim Re_L^{1/2} Re_S^{-1}.
    \label{eqn:cfs_buoy}
\end{equation}
With the assumption (\ref{eqn:cfl_Prandtl_Blasius}), we see that (\ref{eqn:tbl_equation_l}) associated with small $Pr$ becomes
\begin{equation}
    Nu \sim Pr^{1/2} C_T Re_T,
    \label{eqn:Nu_shear_lowPr}
\end{equation}
\noindent and (\ref{eqn:tbl_equation_u}) associated with large $Pr$ becomes
\begin{equation}
    Nu \sim Pr^{1/3} C_T Re_T.
    \label{eqn:Nu_shear_highPr}
\end{equation}
For the limiting case of $Re_L=0$ one can see that relations (\ref{eqn:Nu_shear_lowPr}) and (\ref{eqn:Nu_shear_highPr}) recover the scaling laws for passive transport in Couette \citep{gsy22} or Poiseuille flow \citep{kay93} where the relation between $C_T$ and $Re_T$ depends on whether the kinetic boundary layer is laminar or turbulent. In the presence of a laminar boundary layer, the trivial scaling, $C_T \sim Re_T^{-1}$, applies. When the boundary layer turns turbulent with increased shear forcing, the relation between $C_T$ and $Re_T$ is given by \citet{pra32a} friction law obtained from the log-law mean velocity profile which states
\begin{equation}
    \sqrt{\frac{2}{C_T}} = \frac{1}{k} \ln{ \left( Re_T  \sqrt{\frac{C_T}{8}} \right) } + B,
    \label{eqn:cft_prandtl}
\end{equation}
\noindent with $k \approx 0.41$ \citep{pir14} being the \citet{kar34} constant and $B\approx 5$ \citep{pir14} indicating the log-law intercept. 

When $Re_L$ is not necessarily small and $Re_S \gg Re_L$, we consider that the passive transport relations of \eqref{eqn:Nu_shear_lowPr} or \eqref{eqn:Nu_shear_highPr} remain relevant, and that the dependence of $C_T(Re_T)$ is unchanged to that described above.
For this case of larger $Re_L$, we assume that the thermal dissipation is dominated by contributions from the bulk, so that the global dissipation relation \eqref{eqn:thermal_diss_srb} can be estimated by \eqref{eqn:tdiss_bulk_l} or \eqref{eqn:tdiss_bulk_u}.
When we compare the passive transport relations with the dissipation estimates, we find that $C_L$ must become independent of $Re_L$, with $C_L\sim Pr^{1/2}$ for low $Pr$ and $C_L\sim Pr^{1/3}$ for high $Pr$.
Furthermore, the behaviour of $Re_L$ in these cases can be revealed by combining the passive transport relations \eqref{eqn:Nu_shear_lowPr} or \eqref{eqn:Nu_shear_highPr} with the boundary-layer estimate for the kinetic dissipation rate \eqref{eqn:kdiss_bl}.
For low $Pr$ this produces $Re_L\sim Ra^{1/2} Pr^{-3/4}$, and for high $Pr$ we get $Re_L\sim Ra^{1/2} Pr^{-5/6}$, which exactly match the Reynolds number scaling relations found in the boundary-layer dominated regime $I$ of the GL-theory for classical Rayleigh--B\'enard convection.

\setlength\tabcolsep{0.2em}
\begin{table}
    \begin{center}
        \def~{\hphantom{0}}
        \begin{tabular}{l c c c c c}
            Regime  & $\epsilon_u$ & $\epsilon_\theta$ & BL ratio & $Nu/Nu_R$ & $Re_L/Re_R$ \\[3pt]
            $I_l$ & BL \eqref{eqn:kdiss_bl_approx} & BL \eqref{eqn:tbl_equation_l} & $\lambda_\theta > \lambda_u$ & $(C_L^2 Re_L)^{-1/8}$ & $(C_L^2 Re_L)^{-1/4}$\\[2pt]
            $I_u$ &  BL \eqref{eqn:kdiss_bl_approx} & BL \eqref{eqn:tbl_equation_u} & $\lambda_\theta < \lambda_u$ & $(C_L^2 Re_L)^{1/12}$ & $(C_L^2 Re_L)^{-1/6}$\\[2pt]
            $II_l$ & bulk \eqref{eqn:kdiss_bulk} & BL \eqref{eqn:tbl_equation_l} & $\lambda_\theta > \lambda_u$ & $(Re_T/Re_L)^{-1/5}$ & $(Re_T/Re_L)^{-2/5}$\\[2pt]
            $II_u$ & bulk \eqref{eqn:kdiss_bulk} & BL \eqref{eqn:tbl_equation_u} & $\lambda_\theta < \lambda_u$ & $(C_L^2 Re_L)^{1/5}(Re_T/Re_L)^{-1/5}$ & $(C_L^2 Re_L)^{1/5}(Re_T/Re_L)^{-2/5}$\\[2pt]
            $III_l$ & BL \eqref{eqn:kdiss_bl_approx} & bulk \eqref{eqn:tdiss_bulk_l} & $\lambda_\theta > \lambda_u$ & $(C_L^2 Re_L)^{-1/3}$ & $(C_L^2 Re_L)^{-1/3}$\\[2pt]
            $III_u$ & BL \eqref{eqn:kdiss_bl_approx} & bulk \eqref{eqn:tdiss_bulk_u} & $\lambda_\theta < \lambda_u$ & $(C_L^2 Re_L)^{1/7}$ & $(C_L^2 Re_L)^{-1/7}$\\[2pt]
            $IV_l$ & bulk \eqref{eqn:kdiss_bulk} & bulk \eqref{eqn:tdiss_bulk_l} & $\lambda_\theta > \lambda_u$ & $(Re_T/Re_L)^{-1/2}$ & $(Re_T/Re_L)^{-1/2}$\\[2pt]
            $IV_u$ & bulk \eqref{eqn:kdiss_bulk} & bulk \eqref{eqn:tdiss_bulk_u} & $\lambda_\theta < \lambda_u$ & $(C_L^2 Re_L)^{1/3}(Re_T/Re_L)^{-1/3}$ & $(C_L^2 Re_L)^{1/9}(Re_T/Re_L)^{-4/9}$\\[2pt]
        \end{tabular}
        \caption{
            Scaling relations for the Nusselt number $Nu$ and LSC Reynolds number $Re_L$ in the buoyancy-dominated regime of sheared Rayleigh--B\'enard convection with a laminar Prandtl--Blasius type boundary layer. The first column indicates the GL regime, the second column indicates the bulk or boundary layer (BL) dominance of the kinetic dissipation rate with the applicable scaling estimate in the parenthesis, and the third column indicates the bulk or boundary layer (BL) dominance of the thermal dissipation rate with the applicable scaling estimate in the parenthesis. The fourth column indicates whether the kinetic boundary layer is thicker than the thermal boundary layer or vice-versa. The fourth and fifth columns indicate the scaling relations for $Nu/Nu_R$ and $Re_L/Re_R$ using the values of $Nu$ and $Re_R$ estimated for the pure Rayleigh--B\'enard system from GL-theory.
        }
        \label{tab:buoy-dom_scaling}
  \end{center}
\end{table}

Now, we focus on the buoyancy-dominated regimes where $Re_S/Re_L \lesssim 1$ and $C_S Re_S^3 \ll (Nu-1) Ra Pr^{-2}$ such that the shear contribution to the kinetic dissipation can be neglected in comparison to the buoyancy contribution. With this restriction, we approximate equation (\ref{eqn:kinetic_diss_srb}) as
\begin{equation}
    \epsilon_u \approx \frac{\nu^3}{H^4} (Nu-1) Ra Pr^{-2},
    \label{eqn:kinetic_diss_srb_approx}
\end{equation}
and equation (\ref{eqn:kdiss_bl}) as
\begin{equation}
    \epsilon_{u,BL} \approx \frac{\nu^3}{H^4} C_L Re_L^3.
    \label{eqn:kdiss_bl_approx}
\end{equation}
Using these approximations, we can provide scaling relations between $Nu/Nu_R$ and $Re/Re_R$ for buoyancy-dominated sheared RB system with $Nu_R$ and $Re_R$ being the Nusselt number and Reynolds number associated with the LSC for the pure RB system.
Following the GL-theory, there are various regimes that can be relevant depending on whether the dissipation rates are dominated by boundary layer or bulk contributions, and whether the thermal boundary layer is thicker than the kinetic boundary layer.
For each of these regimes, we combine \eqref{eqn:kinetic_bl} and \eqref{eqn:kinetic_diss_srb_approx} with the relevant estimates for the dominant dissipation rate contribution to give expressions for $Nu(Ra,Pr,C_L, Re_L, Re_T)$ and $Re_L(Ra,Pr,C_L,Re_L, Re_T)$.
Since our approach is consistent with the GL-theory, the $Ra$ and $Pr$ dependence simply recovers the scaling relations $Nu_R(Ra,Pr)$ and $Re_R(Ra,Pr)$ found for the various regimes of pure RB, providing us with scaling relations for $Nu/Nu_R$ and $Re_L/Re_R$ that only depend on $C_L$, $Re_L$, and $Re_T$.
In table \ref{tab:buoy-dom_scaling}, we outline the relevant estimates for the dissipation rates and present the resulting scaling relations. Is it important to note that these scaling relations are only applicable to buoyancy-dominated sheared RB flows with scaling-wise laminar Prandtl--Blasius type boundary layers.

In the buoyancy-dominated classical GL regimes, we can consider the Prandtl--Blasius scaling (\ref{eqn:cfl_Prandtl_Blasius}) to hold for small shear forcing (i.e., $Re_S \lesssim Re_L$). With this assumption, $C_L^2 Re_L \approx 1$, so the values of $Nu$ and $Re_L$ remain unchanged for buoyancy-dominated regimes $I$ and $III$, whereas for buoyancy-dominated regimes $II$ and $IV$, the non-monotonic behaviour of $Nu$ with increasing $Re_S$ becomes apparent. Although $Nu$ seems to decrease with increasing $Re_S$ in the buoyancy-dominant $II$ and $IV$ regimes, it is important to note that this behaviour is subject to the condition that the boundary layer is a laminar one of the Prandtl--Blasius type. If the boundary layer becomes turbulent, the expected decrease in $Nu$ in the buoyancy-dominated regime might disappear. In this study, we will explore the $Nu$ response in the buoyancy-dominated $II_u$ regime where we can still observe the decrease in $Nu$ with increasing $Re_S$ within reasonable computational cost.

\section{Results from the direct numerical simulations} 
\label{sec:numerical}

\subsection{Scheme and procedure}

In this section, we will compare the scaling relations derived in the previous section against the results from our DNS. Equations (\ref{eqn:Navier}) and (\ref{eqn:temp}) are solved numerically using the in-house open-source code ``AFiD'', which is based on a second-order finite-difference scheme \citep{poe15c}. The code has been extensively validated \citep{ver96,ver97,ste10,ste11,koo18}. We impose periodic boundary conditions in the horizontal directions and no-slip boundary conditions at the top and bottom walls. For most simulations, we use domains of aspect ratios $\Gamma_x=8$ and $\Gamma_y=4$. We also performed CRB simulations with $\Gamma_x=48$ and $\Gamma_y=24$ for $Ra=10^7$, $Pr=1$ to study large-scale flow structures. Due to the need for high resolution at large $Ra$, the RB simulations for $Ra=10^{10}$, $Pr=1$ and $Ra=10^{11}$, $Pr=1$ were performed in domains of aspect ratios $\Gamma_x=\Gamma_y=4$, while the RB simulation for $Ra=10^{12}$, $Pr=1$ was performed in domain of aspect ratios $\Gamma_x=\Gamma_y=2$. For the CRB simulations, the wall-velocities $(-U_w)$ and $U_w$ were imposed as Dirichlet boundary conditions on the bottom and top walls, respectively. This is done for numerical reasons \citep{ber13} and does not affect the analysis of the results. The equivalent velocity fields of the CRB system with the bottom wall at rest and the top wall at $2U_w$ can be obtained by a simple Galilean transformation, i.e. by adding $U_w$ to the numerically obtained flow-field. For the PRB simulations, the volume forcing term $\Pi$ is computed at each time-step to ensure a constant mass flow rate \citep{qua16}. 

We use a uniform discretization in the horizontal, periodic directions and a non-uniform grid in the wall-normal direction, in which we employ higher grid resolution in the boundary layers next to the walls. The thermal boundary layer was ensured to be sufficiently resolved according to the resolution requirements put forward by \citet{shi10}. The near-wall resolution is comparable to that of \citet{loz14,pir14,lee18} to ensure that the kinetic boundary layer is sufficiently resolved. The simulations were run for a long enough physical time for the standard deviation of $Nu$ to converge within about 1\% of its mean value. In our previous work \citep{gsy22a}, we verified that the $Nu$ and $Re_{\tau}$ obtained from a domain with $\Gamma_x=8$ and $\Gamma_y=4$ shows a difference of less than 1\% from the $Nu$ obtained from a domain with $\Gamma_x=48$ and $\Gamma_y=24$. This observation is also supported by the fact that $Nu$ for the RB system converges at an approximate aspect ratio $\Gamma_x=\Gamma_y=4$ \citep{ste18}.

\begin{figure}
    \centering
    \includegraphics[width=\textwidth]{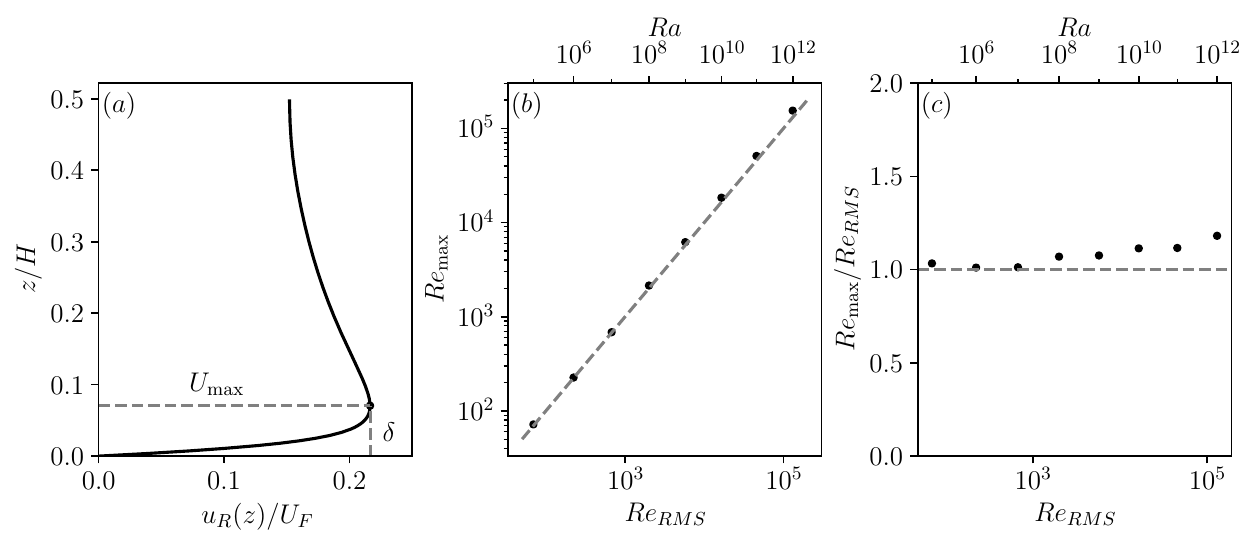}
    \caption{(a) The RMS horizontal velocity profile $u_R(z)$ normalized with the free fall velocity for RB system with $Ra=10^7$, $Pr=1.0$ is shown as a function of wall-normal height $z$, indicating the maximum value $U_{\mathrm{max}}$ occurring at a wall-normal height $\delta$. (b) $Re_{\mathrm{max}}$ obtained from $U_{\mathrm{max}}$ is plotted against the globally averaged RMS velocity $Re_R$ of the RB system. The dashed grey line indicates $Re_{\mathrm{max}} = Re_{RMS}$, showing that these two estimates are virtually identical. The values of $Ra$ are shown on the top for reference. This figure is also available as an \href{https://cocalc.com/share/public_paths/2877f8164894430bc02feb48d07f9834bd2d88bb}{interactive JFM Notebook}.}
    \label{fig:rb_explain}
\end{figure}

Since many of the scaling relations in \S\ref{subsec:regimes} rely on the value of the wind Reynolds number $Re_R$, obtaining an estimate for $Re_R$ from the numerical simulations of pure RB is necessary. The value of $Re_R$ can be estimated in two possible ways. The first estimate can be obtained by using ${Re_\mathrm{max} = U_\mathrm{max}H/\nu}$ where $U_\mathrm{max}$ is the maximum value of the root mean squared (RMS) horizontal velocity profile ${u_R(z) \equiv \sqrt{\left<u_x^2 + u_y^2\right>_{A,t}}}$ at a height $z = \delta$ as shown in figure \ref{fig:rb_explain}a. The second estimate can be obtained by using the global RMS velocity $Re_{RMS} = U_{RMS}H/\nu$ with $U_{RMS} \equiv \sqrt{\left<u_x^2 + u_y^2 + u_z^2\right>_{V,t}}$.
In figure \ref{fig:rb_explain}b, we can see that these estimates are almost identical, providing strong evidence that the LSC driving the wind at the wall also provides the dominant contribution to the mean kinetic energy in RB convection.
In all the results discussed henceforth, we adopt $Re_{RMS}$ as an estimate for $Re_R$.
Unlike in Couette or Poiseuille flow, the mean shear stress at the wall is zero in RB convection.
However, we can use the mean gradient of the RMS horizontal velocity $\left\langle\partial_z u_R(z)\right\rangle_{W,t}$ to calculate the friction coefficient $C_R$ associated with the large-scale circulation. Here $\left\langle ... \right\rangle_{W,t}$ indicates time averaging over the surface of the walls.
The variation of $C_R$ with $Re_R$ is discussed separately in \S\ref{subsec:cf_rb}.

\subsection{Global response parameters}
\label{subsec:nu_retau_results}

\begin{figure}
    \centering
    \includegraphics[width=\textwidth]{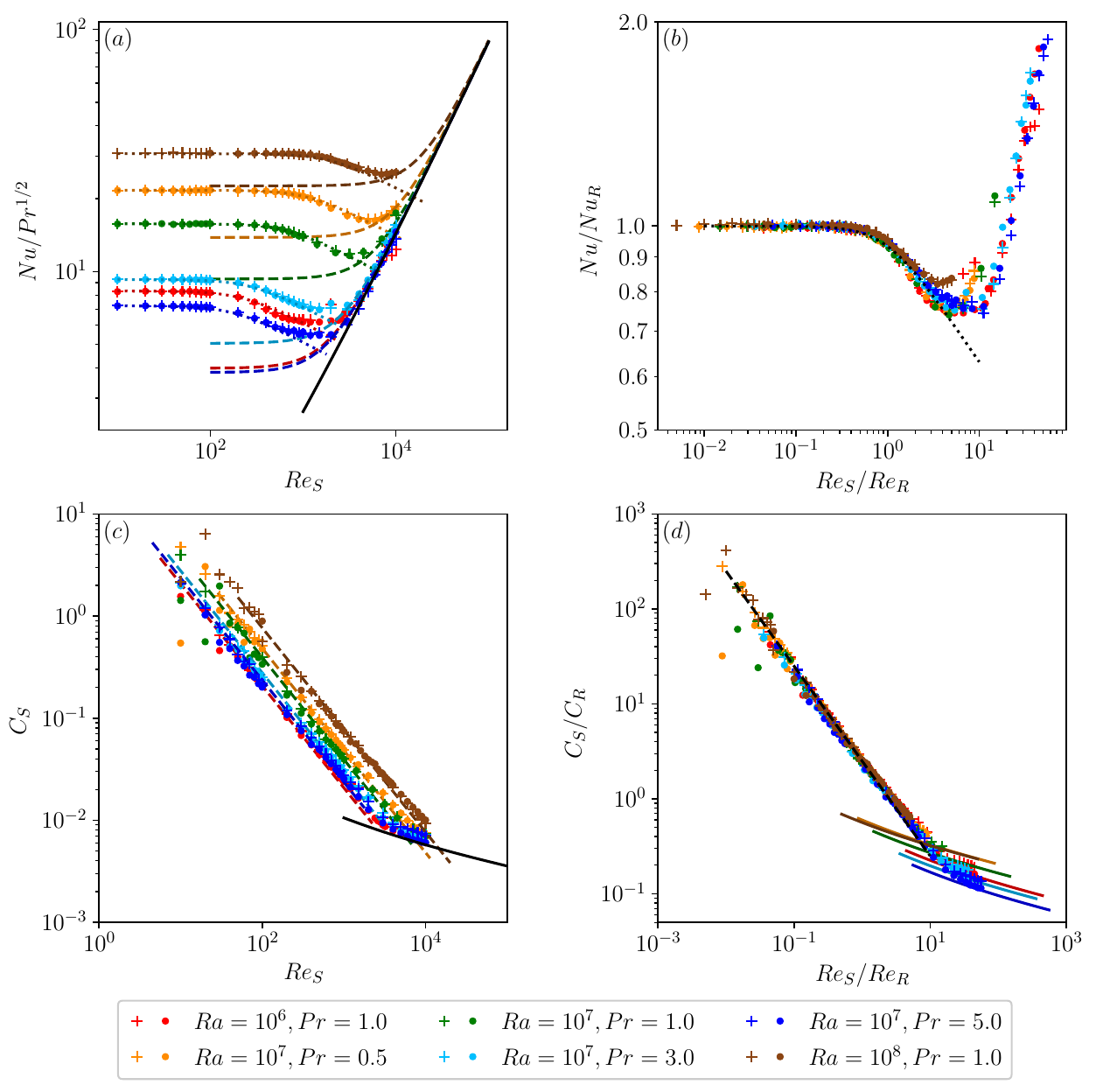}
    \caption{(a) $Nu/Pr^{1/2}$ plotted against $Re_S$. The black solid line indicates (\ref{eqn:nu_shear_trans}), the dashed lines indicate (\ref{eqn:Nu_shear_lowPr}), and the dotted lines indicate (\ref{eqn:nu_buoy_trans}). (b) $Nu/Nu_R$ plotted against $Re_S/Re_R$. Dotted lines indicate (\ref{eqn:nu_buoy_trans}). (c) $C_S$ plotted against $Re_S$. The black solid line indicates (\ref{eqn:cfs_prandtl}) and coloured dashed lines indicate (\ref{eqn:cfs_buoy}). (d) $C_S$ normalised with $C_R$ plotted against $Re_S/Re_R$. Coloured dashed lines indicate (\ref{eqn:cfs_prandtl}) and black dashed line indicates (\ref{eqn:cfs_buoy}). The data for PRB are indicated with plus markers and the data for CRB are indicated with dot markers. This figure is also available as an \href{https://cocalc.com/share/public_paths/528da7b7e647c63e82ad4e71e67f6297944e466c}{interactive JFM Notebook}.}
    \label{fig:nu_cfs}
\end{figure}

We now validate the scaling relations for $Nu$ and $C_S$ derived in \S\ref{subsec:regimes}. Within the parameter range of $Ra$, $Pr$, and $Re$ simulated, we already observe multiple transitions. As shear forcing is increased, we undergo transition from the buoyancy-dominated regime to the shear-dominated regime. For low shear forcing, $\epsilon_u \sim \epsilon_{u,bulk}$ with $\lambda_\theta < \lambda_u$ whereas for high shear forcing, $\epsilon_u \sim \epsilon_{u,BL}$ with $\lambda_\theta/\lambda_u \sim Pr^{1/2}$ \citep{gsy22}. Additionally, at low shear forcing, we observe a laminar Prandtl--Blasius type boundary layer which undergoes a transition into a turbulent one at high shear.

In the buoyancy-dominated regime, we assume the presence of Prandtl--Blasius type kinetic boundary layer with the friction coefficient $C_L$ given by (\ref{eqn:cfl_Prandtl_Blasius}).
For the parameter range of our simulations, the relevant convection regime is $II_u$, so combining the relevant relation from table \ref{tab:buoy-dom_scaling} with \eqref{eqn:cfl_Prandtl_Blasius}, we arrive at
\begin{equation}
    \frac{Nu}{Nu_R} \sim \left(\frac{Re_T}{Re_L}\right)^{-1/5}.
    \label{eqn:IIu_Nuss}
\end{equation}
In the buoyancy-dominated regimes, we also take (\ref{eqn:cfs_buoy}) for the friction coefficient $C_S$ associated with the imposed shear. In order to further simplify these equations, we approximate $Re_L \approx Re_R$ in the buoyancy-dominated regime, which is justified by the weak variation of $Re_L$ with increasing $Re_S$ in (table \ref{tab:buoy-dom_scaling}).
With this assumption, we can rewrite (\ref{eqn:cfs_buoy}) as 
\begin{equation}
    C_S \sim \sqrt{Re_R}/Re_S,
    \label{eqn:cfs_buoy_trans}
\end{equation}
\noindent and use (\ref{eqn:Re_T_approx}) to rewrite \eqref{eqn:IIu_Nuss} as 
\begin{equation}
    Nu/Nu_R \sim \left( \sqrt{1+\left(Re_S^2/Re_R^2\right)} \right)^{-1/5}.
    \label{eqn:nu_buoy_trans}
\end{equation}
\noindent These equations show good agreement with the numerical data plotted in figures \ref{fig:nu_cfs}a-\ref{fig:nu_cfs}d for the buoyancy-dominated regime. Note that equations 3.1 and 3.3 do not explicitly state the dependence of $Nu_R(Ra, Pr)$ or $Re_R(Ra, Pr)$. There is no pure scaling exponent for $Nu_R(Ra)$ in pure RB for regime $II_u$. The values of $Nu_R$ and $Re_R$ used for figures \ref{fig:nu_cfs}b and \ref{fig:nu_cfs}d are obtained from numerical simulations of pure RB. The present entension to the GL-theory assumes that the values of $Nu_R$ and $Re_R$ are known a-priori, and only attemts to provide scaling relations for the normalised quantities $Nu/Nu_R (Re_S/Re_R)$ and $C_S/C_R (Re_S/Re_R)$ in the buoyancy-dominated regime.

\begin{figure}
    \centering
    \includegraphics[width=\textwidth]{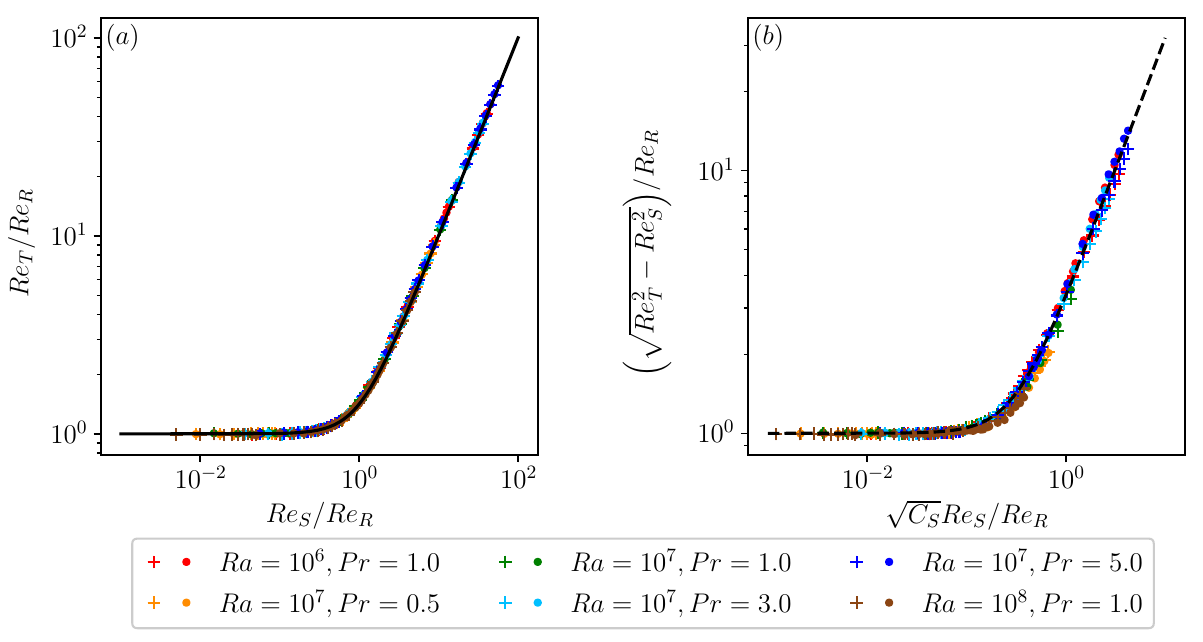}
    \caption{(a) $Re_T/Re_R$ plotted against $Re_S/Re_R$. The black solid line indicates the relation (\ref{eqn:Re_T_approx}). (b) $\left(\sqrt{Re_T^2 - Re_S^2}\right)/Re_R$ plotted against $\sqrt{C_S} Re_S/Re_R$. The black dashed line indicates the relation (\ref{eqn:Re_T}) with $\gamma \approx 10.24$. The data for PRB are indicated with plus markers and the data for CRB are indicated with dot markers. This figure is also available as an \href{https://cocalc.com/share/public_paths/3683630e59e460fcfbad07787d6c94817a4707da}{interactive JFM Notebook}.}
    \label{fig:re_rms}
\end{figure}

For the shear-dominated regime, we observe that the boundary layer becomes turbulent. In this case, (\ref{eqn:cfs_buoy}) is no longer valid. Instead, the relation between $C_T$ and $Re_T$ is given by (\ref{eqn:cft_prandtl}). Note that in the limiting case of very high shear forcing $Re_T \approx Re_S$. Equation (\ref{eqn:cft_prandtl}) can be rewritten as 
\begin{equation}
    \sqrt{\frac{2}{C_S}} = \frac{1}{k} \ln{ \left( Re_S  \sqrt{\frac{C_S}{8}} \right) } + B,
    \label{eqn:cfs_prandtl}
\end{equation}
\noindent and (\ref{eqn:Nu_shear_lowPr}) can be rewritten using (\ref{eqn:kinetic_bl}) as
\begin{equation}
    Nu \sim Pr^{1/2} C_S Re_S
    \label{eqn:nu_shear_trans}
\end{equation}
\noindent which agrees well with the numerical data points in figures \ref{fig:nu_cfs}c and \ref{fig:nu_cfs}d at very high shear forcing. However, it is more useful to substitute the value of $C_T$ obtained from (\ref{eqn:cft_prandtl}) into (\ref{eqn:Nu_shear_lowPr}) and approximate $Re_T$ from (\ref{eqn:Re_T_approx}) as 
\begin{equation}
    Re_T \approx \sqrt{Re_R^2 + Re_S^2}
    \label{eqn:Re_T_buoy_trans}
\end{equation}
to obtain the dashed lines plotted in figure \ref{fig:nu_cfs}a, which show better agreement for a larger range of $Re_S/Re_R$ in the shear-dominated regime. The approximation given by (\ref{eqn:Re_T_buoy_trans}) is then validated in figure \ref{fig:re_rms}a. The additional energy term $\gamma C_S Re_S^2$ from equation (\ref{eqn:Re_T}) that corresponds to the turbulent fluctuations arising from shear forcing is shown in figure \ref{fig:re_rms}b with the value of the prefactor $\gamma \approx 10.24$. Note that the contribution from the fluctuations is much smaller than the contribution from the mean streamwise velocity, thereby making (\ref{eqn:Re_T_buoy_trans}) a good approximation.

\subsection{Large scale circulation}
\label{subsec:lsc_results}

\begin{figure}
    \centering
    \includegraphics[width=\textwidth]{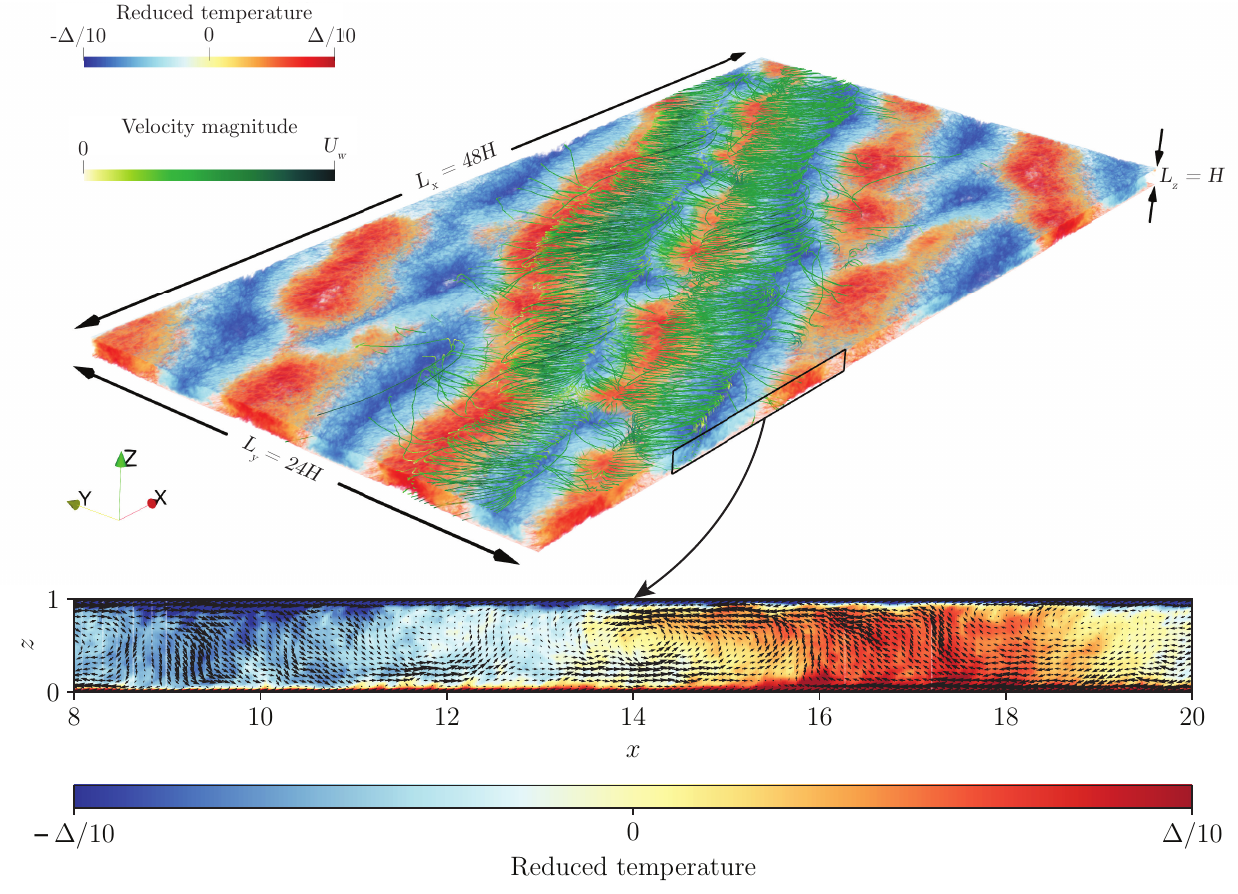}
    \caption{3-D volume visualisation of the time-averaged reduced temperature field of a CRB system with $Ra=10^7$, $Pr=1.0$, $Re_w=1414$, $\Gamma_x=48$, $\Gamma_y=24$ with the averaging time being 100 free-fall time units. The green curves indicate the streamlines of the time-averaged velocity field $\left<\boldsymbol{u}\right>_{t} - U_w \hat{x} $. The spanwise reorientation and streamwise sweeping of the plumes are evident. The zoomed inset shows the time averaged 2D visualisation of the temperature and velocity vectors in the $x-z$ plane. The plumes carry streamwise momentum along with temperature and are swept in the streamwise direction, causing a reduction in the heat transport.}
    \label{fig:lsc_visualisation}
\end{figure}

\begin{figure}
    \centering
    \includegraphics[width=\textwidth]{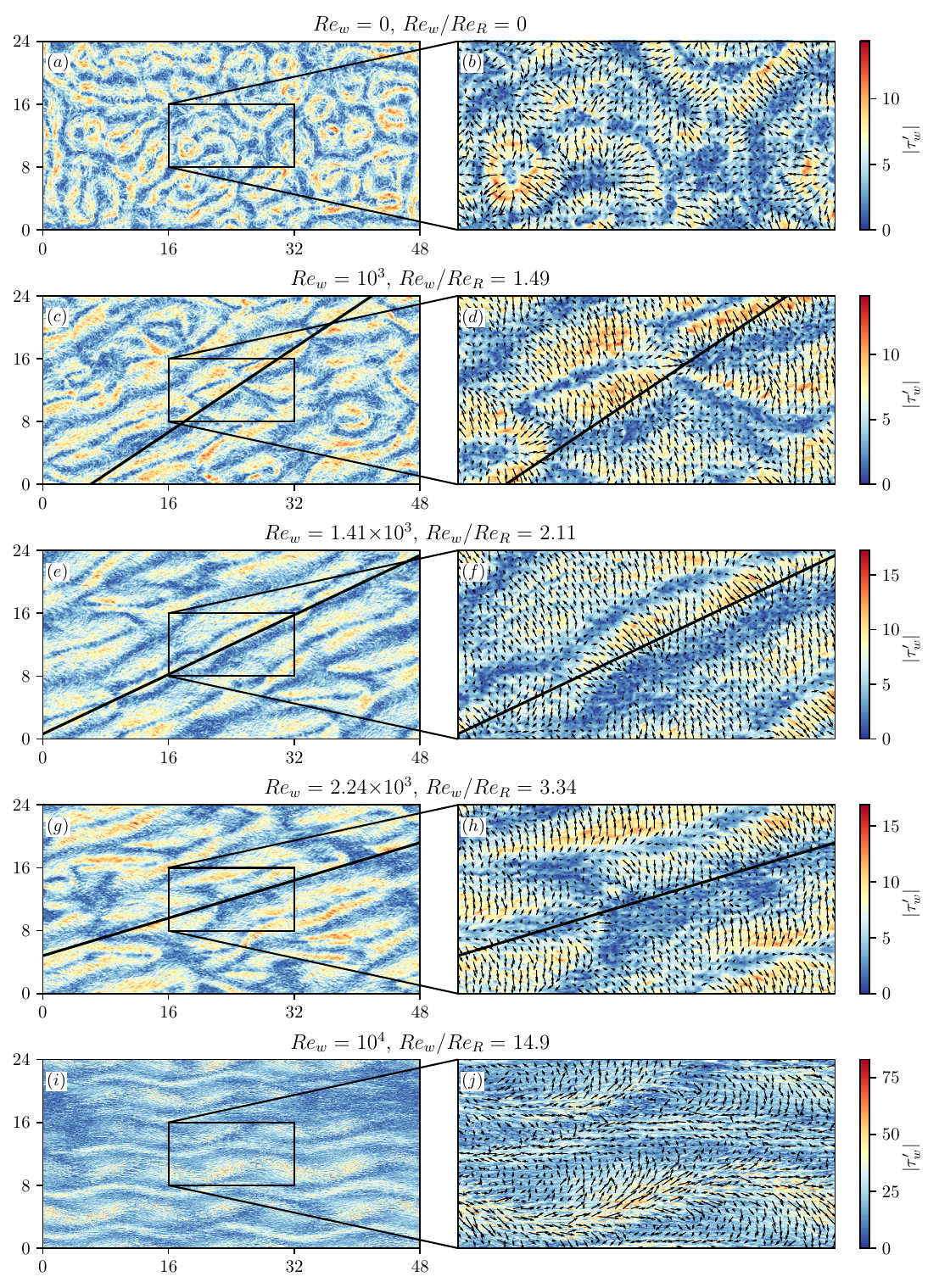}
    \caption{Visualisation of $\boldsymbol{\tau_w^{\prime}}$ for the CRB system with $Ra=10^7$ and $Pr=1$. Plots (c), (e), and (g) are in the buoyancy-dominated regime and plot (i) is in the shear-dominated regime. The colour bars indicate the magnitude of $\boldsymbol{\tau_w^{\prime}}$ while the black arrows in the magnified panels (b), (d), (f), (h) and (j) indicate the direction of $\boldsymbol{\tau_w^{\prime}}$. The black lines in (c-h) indicate the slope of $\bigl(Re_w$/$Re_R\bigr)^{-1}$. This figure is also available as an \href{https://cocalc.com/share/public_paths/4cf976fceffd09de11db96ba6dd3e60cab120eb9}{interactive JFM Notebook}.}
    \label{fig:shear}
\end{figure}

Next, we confirm the theoretical assumptions on the LSC made in \S\ref{subsec:lsc}. The 3-D volume visualisation in figure \ref{fig:lsc_visualisation} shows the streamlines associated with these LSC rolls and it can be seen that they are predominantly oriented in the spanwise direction. However, the large-scale temperature structures in figure \ref{fig:lsc_visualisation} are seen to be aligned neither fully along the streamwise direction, nor fully along the spanwise direction but along a diagonal. The thermal plumes that comprise these large-scale structures experience the advective effects of both $U_L$ in the spanwise direction and $U_S$ in the streamwise direction. Therefore, the orientation of these large scale temperature flow structures in the $x-y$ plane is tilted along a diagonal whose slope is approximately given by $U_L/U_S$. This is made clearer in figure \ref{fig:shear} through the visualisation of the non-dimensional time-averaged local shear stress $\boldsymbol{\tau_w^{\prime}} \equiv (\tau_x^{\prime},\tau_y^{\prime},0)$ computed at the bottom wall in the large aspect ratio CRB system by subtracting the wall-averaged streamwise shear stress in the following way:
\begin{align}
    \tau_x^{\prime} &= H U_{F}^{-1}\left< \partial_z u_x - \left< \partial_z u_x \right>_{x,y,t}\right>_{t}, &
    \tau_y^{\prime} &= H U_{F}^{-1}\left< \partial_z u_y \right>_{t}.
\end{align}

Figure \ref{fig:shear}a shows the wall-shear for the RB system i.e. for $Re_w=0$ and the zoomed inset in figure \ref{fig:shear}b shows the vectors of $\boldsymbol{\tau_w^{\prime}}$. As expected, the LSC rolls are randomly oriented and no global alignment of $\boldsymbol{\tau_w^{\prime}}$ is observed. A visual inspection of figures \ref{fig:shear}c,\ref{fig:shear}e and \ref{fig:shear}g reveals that the large scale flow structures seem to be oriented along the diagonal whose slope is approximately given by $Re_R$/$Re_w$. However, figures \ref{fig:shear}d, \ref{fig:shear}f and \ref{fig:shear}h show that $\boldsymbol{\tau_w^{\prime}}$ is primarily oriented along the spanwise direction in the transitional regime. Figure \ref{fig:shear}i shows the breakdown of the LSC and the formation of large meandering flow-structures \citep{hut07,bla20,bla21b} in the shear-dominated regime, while figure \ref{fig:shear}j shows that $\boldsymbol{\tau_w^{\prime}}$ is predominantly aligned in the streamwise direction in the shear-dominated regime.

\begin{figure}
    \centering
    \includegraphics[width=\textwidth]{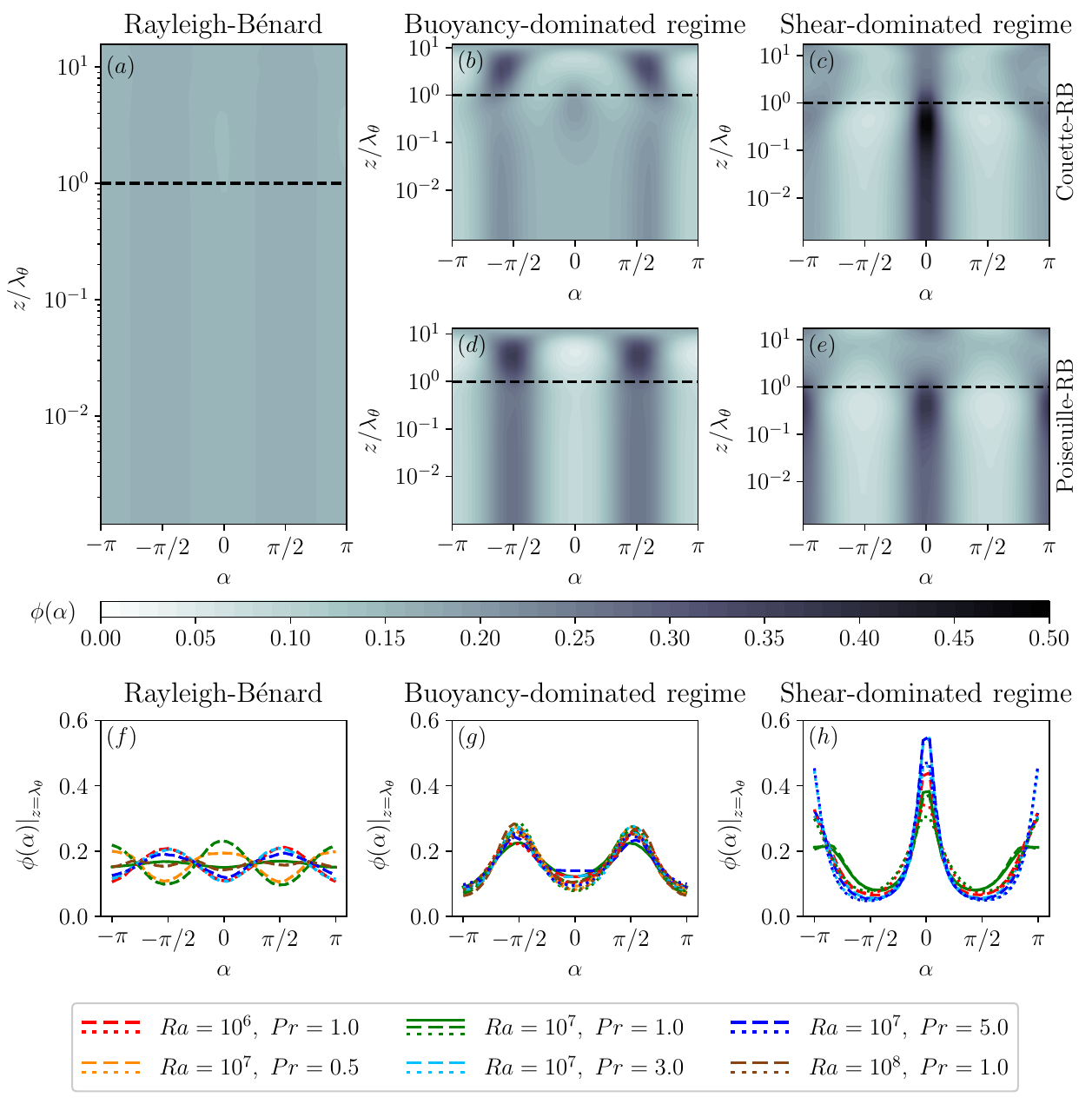}
    \caption{The probability distribution $\phi(\alpha)$ of the flow orientation angle $\alpha$ for all heights $0 \leq z/\lambda_{\theta} \leq Nu$ for $Ra=10^7$, $Pr=1$, $\Gamma_x=48$ and $\Gamma_y=24$ in (a) RB flow (i.e., with $Re_S=0)$, (b,d) buoyancy-dominated regime with $Re_S/Re_R\approx2.11$, and (c,e) shear-dominated regime with $Re_S/Re_R\geq10$. Plots (b,c) are for the CRB system while the plots (d,e) are for the PRB system. The black dashed lines indicate the height of the thermal boundary layer and the colour bar indicates the magnitude of the probability. The probability distribution $\phi(\alpha)$ at the thermal boundary layer height plotted against angle $\alpha$ for various $Ra$ and $Pr$ in (f) RB flow, (g) buoyancy-dominated regime, and (h) shear-dominated regime. The solid lines are the CRB system with $\Gamma_x=48$ and $\Gamma_y=24$, the dashed lines are for the CRB system with $\Gamma_x=8$ and $\Gamma_y=4$, and the dotted lines are for the PRB system with $\Gamma_x=8$ and $\Gamma_y=4$. This figure is also available as an \href{https://cocalc.com/share/public_paths/22200631a3b1f59e8c56f717665fc4109907b55a}{interactive JFM Notebook}.}
    \label{fig:pdf}
\end{figure}

For further confirmation of the changes in the LSC rolls, we study the probability distribution function $\phi(\alpha)$ of the angle $\alpha$ spanned by the horizontal velocity component fluctuations $ \boldsymbol{u_h^\prime} \equiv (u_x^{\prime}, u_y^{\prime}, 0)$ with the streamwise direction $x$. In figures \ref{fig:pdf}a-e it can be seen that the behaviour of $\phi(\alpha)$ is qualitatively quite similar for the CRB and PRB systems. In the RB system, the LSC rolls are randomly oriented as shown by a uniform $\phi(\alpha)$ in figure \ref{fig:pdf}a. In the buoyancy-dominated regime, the LSC roll has a strong tendency to align in the spanwise directions in the transitional regime as shown in \ref{fig:pdf}b,d. In the shear-dominated regime, the velocity fluctuations are predominantly aligned in the streamwise direction as in the case of turbulent Couette/Poiseuille flows as seen in figures \ref{fig:pdf}c,e. At the thermal boundary layer height, the symmetry of $\phi(\alpha)$ about $\alpha=\pm \pi/2$ is strongly suggested for all three regimes by the data from the numerical simulations as shown in figure \ref{fig:pdf}f-h, confirming the assumption made in (\ref{eqn:total_velocity_pdf}). In figure \ref{fig:pdf}f, the small non-uniformity in $\phi(\alpha)$  is attributed to the numerical confinement experienced by the flow structures in domains of a smaller aspect ratio of $\Gamma_x=8$ and $\Gamma_y=4$. For the relatively unconfined case with $\Gamma_x=48$ and $\Gamma_y=24$, the probability distribution is nearly uniform for all values of $\alpha$.

\subsection{Dissipation rates}
\label{subsec:diss_results}

\begin{figure}
    \centering
    \includegraphics[width=\textwidth]{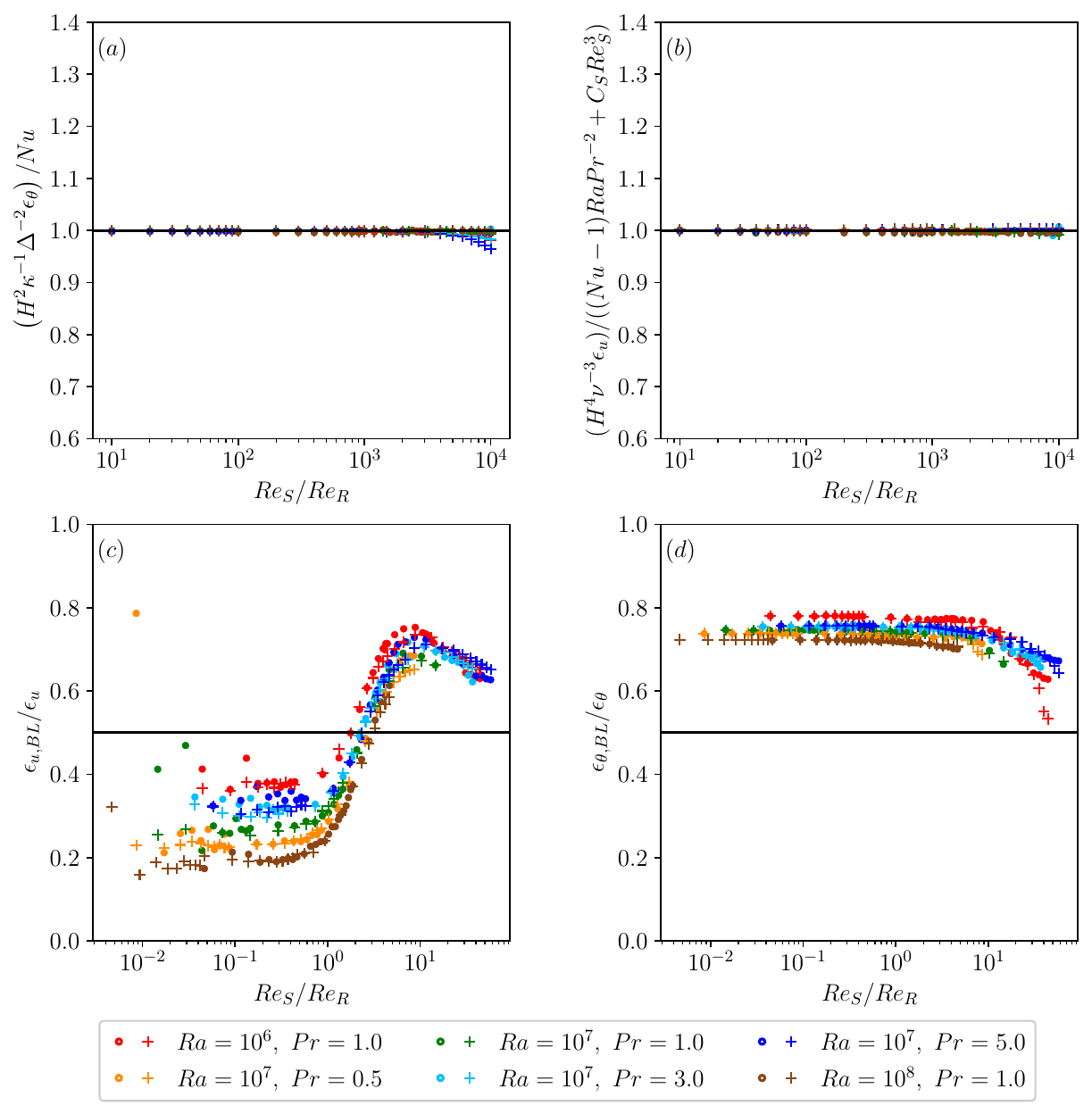}
    \caption{(a) Ratio $ \bigl(H^2 \epsilon_\theta \kappa^{-1} \Delta^2 \bigr)/Nu$ showing good agreement with equation \ref{eqn:thermal_diss_srb} indicated using the solid black line, and (b) ratio $\bigl( H^4 \epsilon_u/ \nu^3 \bigr) / \bigl( (Nu - 1) Ra Pr^{-2} + C_S Re_S^3\bigr)$ plotted against $Re_S$ showing good agreement with equation \ref{eqn:kinetic_diss_srb} of the manuscript shown using the solid black line. (c) Boundary layer contribution to the global kinetic dissipation rate $\epsilon_{u,BL}$ and (d) the boundary layer contribution to the global thermal dissipation rate $\epsilon_{\theta,BL}$. The circle markers are for the CRB system, while plus markers are for the PRB system. This figure is also available as an \href{https://cocalc.com/share/public_paths/510719821511d01ac13aa8c8ea18be7d13019f53}{interactive JFM Notebook}.}
    \label{fig:diss}
\end{figure}

We now investigate the bulk and boundary layer contributions to the global kinetic and thermal dissipation rates described in \S\ref{subsec:diss}. We validate the rigiorous relations given by (\ref{eqn:thermal_diss_srb}) and (\ref{eqn:kinetic_diss_srb}) using the data obtained from numerical simulations as shown in \ref{fig:diss}a and figure \ref{fig:diss}b, respectively. For the range of $Ra$ and $Pr$ studied in this work, as long as the flow is buoyancy-dominated, the system is in the $II_u$ regime with $\epsilon_u \sim \epsilon_{u,bulk}$. With increasing shear forcing, the shear term of (\ref{eqn:kinetic_diss_srb}) increases the boundary layer contribution of the kinetic dissipation due to the formation of streamwise velocity gradients close to the wall. For sufficiently strong shear, the kinetic dissipation will be dominated by the boundary layer contribution with $\epsilon_u \sim \epsilon_{u,BL}$ as shown in \ref{fig:diss}c. On the contrary, figure \ref{fig:diss}d shows that the thermal dissipation rate is dominated by the boundary layer contribution for the entire range of $Re_S$ considered in this study but the contribution reduces noticeably towards higher $Re_S$. For extremely strong shear, a possibility of a transition towards bulk dominance in thermal dissipation cannot be ruled out.

\subsection{Friction coefficient in Rayleigh--B\'enard flow}
\label{subsec:cf_rb}

\begin{figure}
    \centering
    \includegraphics[width=\textwidth]{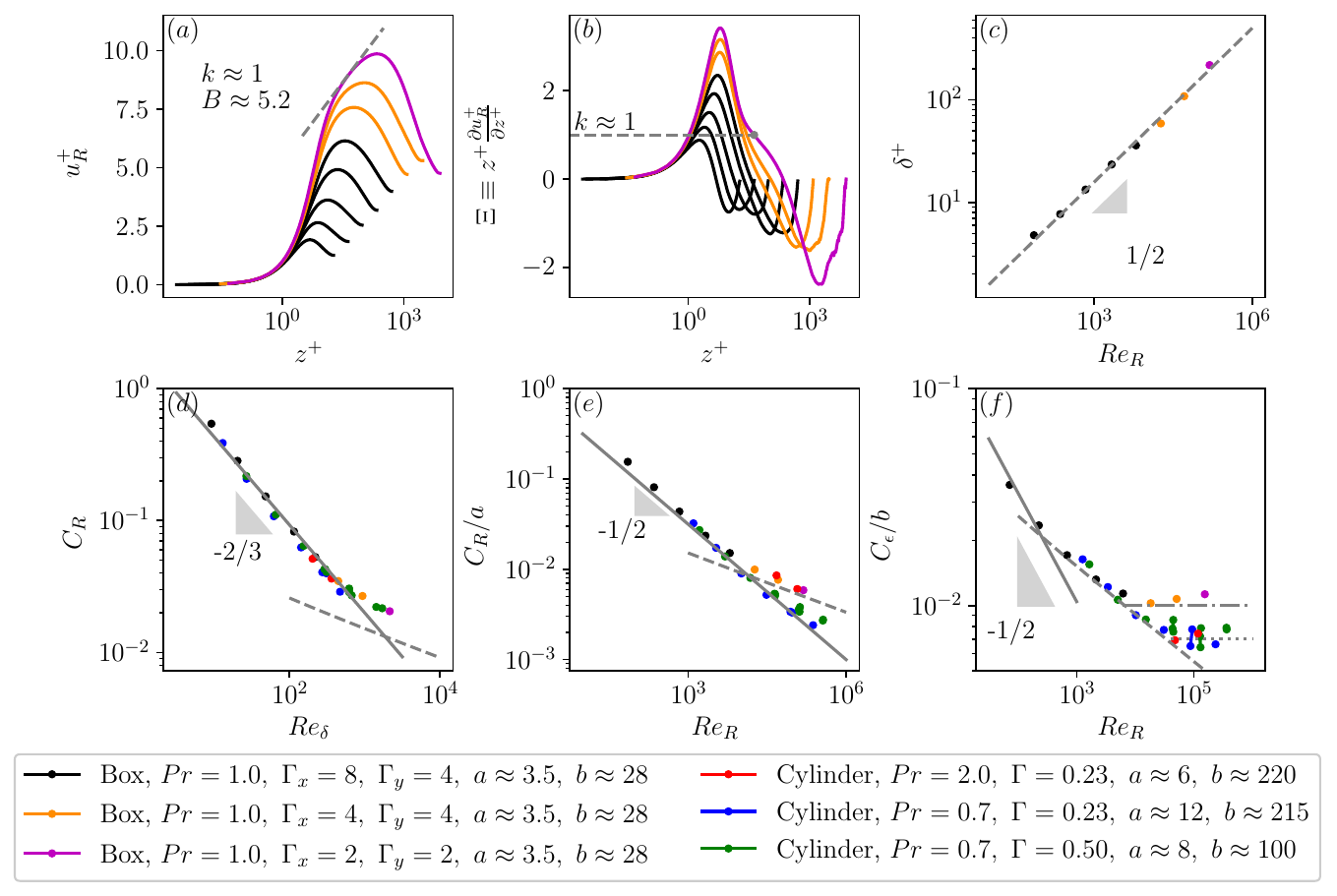}
    \caption{(a) $u_R^+$ plotted against $z^+$, showing the onset of a log-layer with the grey dashed line. The velocity profile for $Ra=10^{12}$, $Pr=1.0$ is indicated in magenta. (b) Diagnostic function plotted against $z^+$ with the inflection point and the corresponding von K\'arm\'an constant indicated. (c) $\delta^+$ plotted against $Re_R$ shows good agreement with $\delta^+ = (1/2) Re_R^{1/2}$, indicated with the dashed grey line. (d) $C_R$ plotted against $Re_\delta$ showing a collapse without any prefactors. The solid grey line indicates $C_R \sim 4Re_\delta^{-2/3}$ and the dashed grey line indicates the modified \citet{pra32a} friction law given by (\ref{eqn:cfr_prandtl}). (e) $C_R/a$ plotted against $Re_R$, showing the Prandtl--Blasius scaling (\ref{eqn:cfl_Prandtl_Blasius}) at low $Re_R$ with the solid grey line and modified \citet{pra32a} friction law given by (\ref{eqn:cfr_prandtl}) at high $Re_R$ with the dashed grey line. (f) $C_\epsilon/b$ given by (\ref{eqn:cfr2}) plotted against $Re_R$. The solid grey line indicates $C_\epsilon \sim Re_R^{-1/2}$ for the regime with $\epsilon_u \sim \epsilon_{u,BL}$, the dashed grey line indicates the modified \citet{pra32a} friction law (\ref{eqn:cfr_prandtl}). At higher $Re_R$, $\epsilon_u \sim \epsilon_{u,bulk}$, leading to $C_\epsilon$ becoming independent of $Re_R$, as indicated with the dash-dotted and dotted grey lines. This figure is also available as an \href{https://cocalc.com/share/public_paths/1eecc949619a06f8d68731e4abe91ab90837eaa0}{interactive JFM Notebook}.}
    \label{fig:cf_rb}
\end{figure}

So far, we have considered only the laminar Prandtl--Blasius type kinetic boundary layers in the buoyancy-dominated regime with the boundary layer only becoming turbulent in highly shear-dominated regime. However, we now consider the possibility of turbulent kinetic boundary layers even in pure RB flow which corresponds to the so-called ``ultimate'' regime in RB flow \citep{kra62,gro11,roc20,loh23}. While the existence of logarithmic temperature profiles has already been observed at high $Ra$ \citep{gro12,ahl12,ahl14}, it is yet to be seen if logarithmic behaviour given by 
\begin{equation}
    u_R^+ (z^+) = \frac{1}{k} \log (z^+) + B
\end{equation}
with,
\begin{align}
    u_R^+ (z^+) &= u_R(z) /u_\tau, & z^+ &= \frac{z u_\tau}{2 \nu}, & u_R(z) &\equiv \sqrt{\left<u_x^2 + u_y^2\right>_{A,t}},  & u_\tau &= \sqrt{\nu \left<\partial_z u_R\right>_{W,t}}. 
\end{align}
can be observed in the velocity profiles. 

At the highest thermal forcing studied in this work with $Ra=10^{12}$, we start to also observe some hints of what could possibly be the onset of a log-layer (see figure \ref{fig:cf_rb}a), although it cannot be conclusively confirmed with the currently available data. Assuming that such a log-layer could exist, an estimate of the modified von K\'arm\'an constant $k$ is obtained from the inflection point of the diagnostic function plotted in figure \ref{fig:cf_rb}b, giving $k \approx 1$. Correspondingly, the intercept $B \approx 5.2$ is found by fitting the data as shown in figure \ref{fig:cf_rb}a.

In the presence of such a logarithmic layer, we can now hypothesise about a relation for $C_R$ which is analogous to relation (\ref{eqn:cfs_prandtl}) for $C_S$ and (\ref{eqn:cft_prandtl}) for $C_T$ by stating that in the limit of highly turbulent boundary layer, $u_R^+=U_R$ at $z^+=\delta^+$, where $\delta$ is the wall-normal distance to the peak velocity shown in figure \ref{fig:re_rms}.
This gives us
\begin{align}
    \sqrt{\frac{2}{C_R}} &= \frac{1}{k} \ln{ \left( Re_\delta  \sqrt{\frac{C_R}{8}} \right) } + B, & Re_\delta = U_R \delta / \nu
     \label{eqn:cfr_prandtl}
\end{align}
If applicable to high $Ra$ RB flow, the general form of this equation could suggest universality in the behaviour of wall-bounded flows even though the values of $k$ and $B$ might be different from those observed for pipe or channel flows. By plotting $\delta^+$ against $Re_R$, we find a very good fit with the scaling $\delta^+ \sim Re_R^{1/2}$ as shown in figure \ref{fig:cf_rb}c. At present, we can only provide this scaling empirically because more data at extremely high $Ra$ are needed to understand the dynamics of the turbulent boundary layer which is difficult due to the high computational expense of such numerical simulations. Plotting $C_R$ against $Re_\delta$ in \ref{fig:cf_rb}d, we find a nice collapse of the data for different aspect ratios and geometries. At low $Re_R$, $C_R$ seems to scale as $Re_\delta^{-2/3}$ which can be obtained from the empirical scaling $\delta^+ \sim Re_R^{1/2}$ shown in figure \ref{fig:cf_rb}c and the assumption of a Prandtl--Blasius type scaling of $C_R \sim Re_R^{-1/2}$. At high $Re_R$, we see good qualitative agreement between the data and the modified \citet{pra32a} friction law given by (\ref{eqn:cfr_prandtl}).

We can further investigate the behaviour of $C_R$ by computing
\begin{subequations}
    \begin{align}
        C_R &\equiv a Re_R^{-1/2}, \label{eqn:cfr1} \\
        C_\epsilon &\equiv b (Nu - 1) Ra Pr^{-2} Re_R^{-3}, & 
        \begin{cases}
            C_\epsilon \sim C_R & \text{ if } \epsilon_u \sim \epsilon_{u,BL} \\ 
            C_\epsilon = \text{constant} & \text{ if } \epsilon_u \sim \epsilon_{u,bulk}
        \end{cases}
        \label{eqn:cfr2}
    \end{align}
\end{subequations}
\noindent where $a$ and $b$ are prefactors that are obtained by fitting the data from figures \ref{fig:cf_rb}e and \ref{fig:cf_rb}f, accounting for the effects of aspect ratio and geometry. Note that when the kinetic dissipation of the RB system is dominated by the contribution from the boundary layer, $C_\epsilon \sim C_R$ in equation (\ref{eqn:cfr2}). The data points at lower $Re_R$ in figure \ref{fig:cf_rb}e are observed to follow (\ref{eqn:cfl_Prandtl_Blasius}) which is consistent with the Prandtl--Blasius scaling. For higher $Re_R$ corresponding to $Ra>10^{10}$, $C_R$ shows better agreement with the \citet{pra32a} friction law given by (\ref{eqn:cfr_prandtl}). Figure \ref{fig:cf_rb}f shows a very similar behaviour to figure \ref{fig:cf_rb}e but, in addition, we observe that $C_\epsilon$ becomes independent of $Re_R$ at very high values of $Re_R$. Although this does not reflect the true dependence of $C_R$ on $Re_R$, this apparent dependence is expected because the kinetic dissipation of the RB system undergoes a transition from being dominated by the boundary layer to being dominated by the bulk \citep{loh94a}. It should also be noted that this transition occurs at higher $Re_R$ for the more confined cylindrical RB simulations because more kinetic driving is required to overcome the viscous dissipation in the additional boundary layers on the side walls that are not present in the periodic box RB simulations \citep{ahl22}.

\section{Conclusions} 
\label{sec:conclusion}

In summary, we have developed a framework by extending the GL-theory for RB turbulence to sheared RB turbulence. As in the case of RB flow, we observe that there are no pure scaling exponents for the Nusselt number $Nu$ and the friction coefficient $C_S$. This also holds for high thermal or shear driving where the boundary layers no longer obey scaling relations associated with the Prandtl--Blasius \citep{pra04,bla08} boundary layer theory but start to become more turbulent. In such cases, we observe that the relation for $C_S(Re_S)$ is well described by the friction law of \citet{pra32a}. In addition, we find that a modified version of the \cites{pra32a} friction law for the convective large-scale circulation $C_L(Re_L)$ analogous to $C_S(Re_S)$ agrees well with the DNS data, suggesting some sort of universality in the relation between the shear stress and the flow velocity that generates that shear. 

It is also interesting to note that the relations are identical for CRB or PRB systems once the appropriate velocity scale is chosen as a control parameter. This suggests that the flow physics is not strongly affected by the geometry of the system or by the way in which shear forcing is applied. The flow characteristics of these systems are essentially determined by the ratio of shear driving to thermal driving, given by $Re_S/Re_R$. Applying shear to the RB system causes increased coherence in the streamwise direction and leads to a re-orientation of the LSC rolls and causes them to align more in the spanwise direction, with the thermal plumes also transporting the momentum imparted by the shear forcing. In the buoyancy-dominated regime with $Re_S \simeq Re_R$, this may lead to enhanced streamwise mixing between hot and cold plumes at a timescale that is smaller than the timescale of heat diffusion at the wall. This leads to heat entrapment in the bulk and a reduction in $Nu$. 

Taking into account the orientation of the LSC rolls and the bulk dominance of $\epsilon_u$ in the buoyancy-dominated regime, we show that the orientation of large-scale flow structures can also be predicted to a reasonable degree by the ratio $Re_S/Re_R$, and we provide scaling relations for the $Nu(Re_S/Re_R)$ and $C_S(Re_S/Re_R)$ which are shown to agree well with the numerical simulations for $10^6 \leq Ra \leq 10^8$, $0.5 \leq Pr \leq 5.0$, and $0 \leq Re_S \leq 10^4$. However, the evidence from the DNS is limited at the moment due to its high computational costs, thereby restricting the parameter range in which the proposed scaling laws can be validated. Simulations for very high or very low values of $Pr$ as well as for high $Ra$ or $Re_S$ can be very demanding, and it remains to be seen if the assumptions made in this work and the extended theory hold well in other control parameter ranges.


\backsection[Acknowledgements]{The authors gratefully acknowledge Robert Hartmann, Alexander Blass, and Marco De Paoli for fruitful discussions.}

\backsection[Funding]{This work was financially supported by the ERC starting grant (2018) for the project “UltimateRB” and the Twente Max-Planck Center. This work was carried out on the Dutch national e-infrastructure with the support of SURF Cooperative. We acknowledge PRACE for awarding us access to MareNostrum at Barcelona Supercomputing Center (BSC), Spain (Project 2020225335 and 2020235589). The authors gratefully acknowledge the Gauss Centre for Supercomputing e.V. (www.gauss-centre.eu) for funding this project by providing computing time on the GCS Supercomputer SuperMUC at Leibniz Supercomputing Centre (www.lrz.de).}

\backsection[Declaration of interests]{The authors report no conflict of interest.}


\backsection[Author ORCIDs]{
G. S. Yerragolam, https://orcid.org/0000-0002-8928-2029; \\
C. J. Howland, https://orcid.org/0000-0003-3686-9253; \\
R. J. A. M. Stevens, https://orcid.org/0000-0001-6976-5704; \\
R. Verzicco, https://orcid.org/0000-0002-2690-9998; \\
O. Shishkina, https://orcid.org/0000-0002-6773-6464; \\
D. Lohse, https://orcid.org/0000-0003-4138-2255}

\backsection[Author contributions]{The theoretical framework was developed by D. Lohse, O. Shishkina, R. J. A. M. Stevens, C. J. Howland, and G. S. Yerragolam. The code "AFiD" for the direct numerical simulations was developed by R. Verzicco while the numerical simulations were performed by G. S. Yerragolam. All authors were involved in the preparation of the manuscript.}

\appendix

\begin{landscape}

\section{Simulation Parameters}\label{appA}

In table \ref{tab:gridcases} we provide the physical and numerical input parameters used for the new sheared RB simulations conducted for this study.
In addition to the new simulations, data for large aspect ratio simulations with $\Gamma_x = 48$ and $\Gamma_y = 24$ is taken from \citet{gsy22a}, and data for high $Ra$ cylindrical simulations is taken from \citet{har23}.

\begin{longtable}[c]{| r | r | r | r | r | r | r | r | r | r | r | r | r | r | r | r | r | r |}
            
\ctr{System} & \ctr{$\Gamma_x$} & \ctr{$\Gamma_y$} & \ctr{$N_x$} & \ctr{$N_y$} & \ctr{$N_z$} & \ctr{$x^+$} & \ctr{$y^+$} & \ctr{$z^+_w$} & \ctr{$z^+_m$} & \ctr{$N_{BL}$} & \ctr{$Ra$}  & \ctr{$Pr$} & \ctr{$Re_S$} & \ctr{$Nu_{w}$} & \ctr{$Nu_{\epsilon_\theta}$} & \ctr{$Nu_{\epsilon_u}$} & \ctr{$C_S$} \\
\hline
\hline

RB & 8 & 4 & 1024 & 512 & 192 & - & - & - & - & 41 & $10^{5}$ & 1.0 & 0 & 4.38 & 4.38 & 4.38 & - \\
RB & 8 & 4 & 1536 & 768 & 256 & - & - & - & - & 40 & $10^{6}$ & 1.0 & 0 & 8.31 & 8.31 & 8.32 & - \\
RB & 8 & 4 & 1536 & 768 & 256 & - & - & - & - & 29 & $10^{7}$ & 0.5 & 0 & 15.32 & 15.33 & 15.32 & - \\
RB & 8 & 4 & 1536 & 768 & 256 & - & - & - & - & 28 & $10^{7}$ & 1.0 & 0 & 15.75 & 14.47 & 15.73 & - \\
RB & 8 & 4 & 2048 & 1024 & 384 & - & - & - & - & 42 & $10^{7}$ & 3.0 & 0 & 16.05 & 16.04 & 16.02 & - \\
RB & 8 & 4 & 2048 & 1024 & 384 & - & - & - & - & 42 & $10^{7}$ & 5.0 & 0 & 16.14 & 16.14 & 16.19 & - \\
RB & 8 & 4 & 2048 & 1024 & 384 & - & - & - & - & 29 & $10^{8}$ & 1.0 & 0 & 30.68 & 30.68 & 30.71 & - \\
RB & 8 & 4 & 3072 & 1536 & 512 & - & - & - & - & 28 & $10^{9}$ & 1.0 & 0 & 61.83 & 61.86 & 61.98 & - \\
RB & 4 & 4 & 2048 & 2048 & 768 & - & - & - & - & 28 & $10^{10}$ & 1.0 & 0 & 129.78 & 129.80 & 129.80 & - \\
RB & 4 & 4 & 3072 & 3072 & 1024 & - & - & - & - & 24 & $10^{11}$ & 1.0 & 0 & 287.97 & 288.02 & 287.26 & - \\
RB & 2 & 2 & 2048 & 2048 & 1536 & - & - & - & - & 23 & $10^{12}$ & 1.0 & 0 & 709.21 & 709.50 & 732.83 & - \\
\hline
& & & & & & & & & & & & & & & & &  \\[-6.5pt]
CRB & 8 & 4 & 1536 & 768 & 256 & 0.05 & 0.05 & 0.00 & 0.05 & 40 & $10^{6}$ & 1.0 & 10 & 8.26 & 8.27 & 8.27 & 1.56 \\
CRB & 8 & 4 & 1536 & 768 & 256 & 0.08 & 0.08 & 0.00 & 0.09 & 40 & $10^{6}$ & 1.0 & 20 & 8.29 & 8.29 & 8.29 & 1.09 \\
CRB & 8 & 4 & 1536 & 768 & 256 & 0.07 & 0.07 & 0.00 & 0.09 & 40 & $10^{6}$ & 1.0 & 30 & 8.26 & 8.26 & 8.25 & 0.459 \\
CRB & 8 & 4 & 1536 & 768 & 256 & 0.11 & 0.11 & 0.00 & 0.12 & 40 & $10^{6}$ & 1.0 & 40 & 8.28 & 8.27 & 8.26 & 0.511 \\
CRB & 8 & 4 & 1536 & 768 & 256 & 0.12 & 0.12 & 0.00 & 0.14 & 40 & $10^{6}$ & 1.0 & 50 & 8.26 & 8.26 & 8.25 & 0.399 \\
CRB & 8 & 4 & 1536 & 768 & 256 & 0.13 & 0.13 & 0.00 & 0.15 & 40 & $10^{6}$ & 1.0 & 60 & 8.29 & 8.29 & 8.29 & 0.327 \\
CRB & 8 & 4 & 1536 & 768 & 256 & 0.14 & 0.14 & 0.00 & 0.17 & 40 & $10^{6}$ & 1.0 & 70 & 8.25 & 8.24 & 8.24 & 0.3 \\
CRB & 8 & 4 & 1536 & 768 & 256 & 0.15 & 0.15 & 0.00 & 0.17 & 40 & $10^{6}$ & 1.0 & 80 & 8.21 & 8.21 & 8.22 & 0.251 \\
CRB & 8 & 4 & 1536 & 768 & 256 & 0.16 & 0.16 & 0.00 & 0.18 & 40 & $10^{6}$ & 1.0 & 90 & 8.19 & 8.19 & 8.18 & 0.224 \\
CRB & 8 & 4 & 1536 & 768 & 256 & 0.17 & 0.17 & 0.00 & 0.19 & 40 & $10^{6}$ & 1.0 & 100 & 8.18 & 8.18 & 8.19 & 0.203 \\
CRB & 8 & 4 & 1536 & 768 & 256 & 0.23 & 0.23 & 0.00 & 0.28 & 41 & $10^{6}$ & 1.0 & 200 & 7.84 & 7.84 & 7.85 & 0.102 \\
CRB & 8 & 4 & 1536 & 768 & 256 & 0.29 & 0.29 & 0.00 & 0.34 & 42 & $10^{6}$ & 1.0 & 300 & 7.47 & 7.47 & 7.47 & 0.0676 \\
CRB & 8 & 4 & 1536 & 768 & 256 & 0.34 & 0.34 & 0.00 & 0.40 & 43 & $10^{6}$ & 1.0 & 400 & 6.97 & 6.97 & 6.95 & 0.0546 \\
CRB & 8 & 4 & 1536 & 768 & 256 & 0.39 & 0.39 & 0.01 & 0.45 & 44 & $10^{6}$ & 1.0 & 500 & 6.66 & 6.67 & 6.66 & 0.0442 \\
CRB & 8 & 4 & 1536 & 768 & 256 & 0.43 & 0.43 & 0.01 & 0.51 & 45 & $10^{6}$ & 1.0 & 600 & 6.46 & 6.46 & 6.45 & 0.038 \\
CRB & 8 & 4 & 1536 & 768 & 256 & 0.47 & 0.47 & 0.01 & 0.55 & 46 & $10^{6}$ & 1.0 & 700 & 6.35 & 6.35 & 6.35 & 0.0332 \\
CRB & 8 & 4 & 1536 & 768 & 256 & 0.51 & 0.51 & 0.01 & 0.60 & 46 & $10^{6}$ & 1.0 & 800 & 6.31 & 6.31 & 6.33 & 0.0296 \\
CRB & 8 & 4 & 1536 & 768 & 256 & 0.54 & 0.54 & 0.01 & 0.64 & 46 & $10^{6}$ & 1.0 & 900 & 6.27 & 6.28 & 6.26 & 0.0268 \\
CRB & 8 & 4 & 1536 & 768 & 256 & 0.55 & 0.55 & 0.01 & 0.64 & 46 & $10^{6}$ & 1.0 & 920 & 6.25 & 6.25 & 6.26 & 0.0261 \\
CRB & 8 & 4 & 1536 & 768 & 256 & 0.55 & 0.55 & 0.01 & 0.65 & 46 & $10^{6}$ & 1.0 & 940 & 6.27 & 6.27 & 6.27 & 0.0255 \\
CRB & 8 & 4 & 1536 & 768 & 256 & 0.56 & 0.56 & 0.01 & 0.66 & 46 & $10^{6}$ & 1.0 & 960 & 6.24 & 6.24 & 6.22 & 0.0252 \\
CRB & 8 & 4 & 1536 & 768 & 256 & 0.57 & 0.57 & 0.01 & 0.66 & 46 & $10^{6}$ & 1.0 & 980 & 6.20 & 6.19 & 6.21 & 0.0245 \\
CRB & 8 & 4 & 1536 & 768 & 256 & 0.57 & 0.57 & 0.01 & 0.67 & 46 & $10^{6}$ & 1.0 & 1000 & 6.21 & 6.21 & 6.21 & 0.0241 \\
CRB & 8 & 4 & 1536 & 768 & 256 & 0.63 & 0.63 & 0.01 & 0.74 & 46 & $10^{6}$ & 1.0 & 1200 & 6.19 & 6.19 & 6.18 & 0.0203 \\
CRB & 8 & 4 & 1536 & 768 & 256 & 0.71 & 0.71 & 0.01 & 0.83 & 46 & $10^{6}$ & 1.0 & 1500 & 6.19 & 6.19 & 6.19 & 0.0164 \\
CRB & 8 & 4 & 1536 & 768 & 256 & 0.82 & 0.82 & 0.01 & 0.97 & 46 & $10^{6}$ & 1.0 & 2000 & 6.26 & 6.26 & 6.25 & 0.0125 \\
CRB & 8 & 4 & 1536 & 768 & 256 & 0.91 & 0.91 & 0.01 & 1.06 & 46 & $10^{6}$ & 1.0 & 2400 & 6.29 & 6.29 & 6.30 & 0.0105 \\
CRB & 8 & 4 & 1536 & 768 & 256 & 0.95 & 0.95 & 0.01 & 1.12 & 45 & $10^{6}$ & 1.0 & 2600 & 6.39 & 6.38 & 6.38 & 0.00988 \\
CRB & 8 & 4 & 1536 & 768 & 256 & 1.01 & 1.01 & 0.01 & 1.18 & 44 & $10^{6}$ & 1.0 & 2800 & 6.72 & 6.72 & 6.73 & 0.00957 \\
CRB & 8 & 4 & 1536 & 768 & 256 & 1.05 & 1.05 & 0.02 & 1.23 & 44 & $10^{6}$ & 1.0 & 3000 & 6.74 & 6.73 & 6.67 & 0.00897 \\
CRB & 8 & 4 & 1536 & 768 & 256 & 1.10 & 1.10 & 0.02 & 1.29 & 44 & $10^{6}$ & 1.0 & 3200 & 6.92 & 6.92 & 6.86 & 0.00867 \\
CRB & 8 & 4 & 1536 & 768 & 256 & 1.31 & 1.31 & 0.02 & 1.54 & 41 & $10^{6}$ & 1.0 & 4000 & 7.83 & 7.83 & 7.82 & 0.0079 \\
CRB & 8 & 4 & 1536 & 768 & 256 & 1.58 & 1.58 & 0.02 & 1.86 & 38 & $10^{6}$ & 1.0 & 5000 & 9.16 & 9.16 & 9.12 & 0.00738 \\
CRB & 8 & 4 & 1536 & 768 & 256 & 1.85 & 1.85 & 0.03 & 2.17 & 35 & $10^{6}$ & 1.0 & 6000 & 10.45 & 10.46 & 10.45 & 0.00702 \\
CRB & 8 & 4 & 1536 & 768 & 256 & 2.08 & 2.08 & 0.03 & 2.44 & 33 & $10^{6}$ & 1.0 & 7000 & 11.51 & 11.50 & 11.51 & 0.00652 \\
CRB & 8 & 4 & 1536 & 768 & 256 & 2.35 & 2.35 & 0.03 & 2.76 & 31 & $10^{6}$ & 1.0 & 8000 & 12.87 & 12.86 & 12.86 & 0.00639 \\
CRB & 8 & 4 & 1536 & 768 & 256 & 2.59 & 2.59 & 0.04 & 3.04 & 30 & $10^{6}$ & 1.0 & 9000 & 13.92 & 13.92 & 13.91 & 0.00613 \\
CRB & 8 & 4 & 1536 & 768 & 256 & 2.86 & 2.86 & 0.04 & 3.35 & 29 & $10^{6}$ & 1.0 & 10000 & 15.18 & 15.17 & 15.17 & 0.00602 \\
\hline
& & & & & & & & & & & & & & & & &  \\[-6.5pt]
CRB & 8 & 4 & 1536 & 768 & 256 & 0.03 & 0.03 & 0.00 & 0.03 & 29 & $10^{7}$ & 0.5 & 10 & 15.30 & 15.30 & 15.30 & 0.543 \\
CRB & 8 & 4 & 1536 & 768 & 256 & 0.13 & 0.13 & 0.00 & 0.15 & 29 & $10^{7}$ & 0.5 & 20 & 15.29 & 15.29 & 15.29 & 3.05 \\
CRB & 8 & 4 & 1536 & 768 & 256 & 0.12 & 0.12 & 0.00 & 0.14 & 29 & $10^{7}$ & 0.5 & 30 & 15.32 & 15.32 & 15.33 & 1.14 \\
CRB & 8 & 4 & 1536 & 768 & 256 & 0.13 & 0.13 & 0.00 & 0.16 & 29 & $10^{7}$ & 0.5 & 40 & 15.26 & 15.26 & 15.26 & 0.829 \\
CRB & 8 & 4 & 1536 & 768 & 256 & 0.17 & 0.17 & 0.00 & 0.20 & 29 & $10^{7}$ & 0.5 & 50 & 15.26 & 15.27 & 15.26 & 0.843 \\
CRB & 8 & 4 & 1536 & 768 & 256 & 0.16 & 0.16 & 0.00 & 0.19 & 29 & $10^{7}$ & 0.5 & 60 & 15.22 & 15.22 & 15.22 & 0.554 \\
CRB & 8 & 4 & 1536 & 768 & 256 & 0.22 & 0.22 & 0.00 & 0.26 & 29 & $10^{7}$ & 0.5 & 70 & 15.31 & 15.31 & 15.33 & 0.746 \\
CRB & 8 & 4 & 1536 & 768 & 256 & 0.22 & 0.22 & 0.00 & 0.26 & 29 & $10^{7}$ & 0.5 & 80 & 15.34 & 15.35 & 15.35 & 0.558 \\
CRB & 8 & 4 & 1536 & 768 & 256 & 0.21 & 0.21 & 0.00 & 0.25 & 29 & $10^{7}$ & 0.5 & 90 & 15.26 & 15.26 & 15.24 & 0.398 \\
CRB & 8 & 4 & 1536 & 768 & 256 & 0.25 & 0.25 & 0.00 & 0.30 & 29 & $10^{7}$ & 0.5 & 100 & 15.35 & 15.36 & 15.36 & 0.477 \\
CRB & 8 & 4 & 1536 & 768 & 256 & 0.35 & 0.35 & 0.01 & 0.41 & 29 & $10^{7}$ & 0.5 & 200 & 15.31 & 15.31 & 15.31 & 0.227 \\
CRB & 8 & 4 & 1536 & 768 & 256 & 0.44 & 0.44 & 0.01 & 0.52 & 29 & $10^{7}$ & 0.5 & 300 & 15.24 & 15.23 & 15.23 & 0.16 \\
CRB & 8 & 4 & 1536 & 768 & 256 & 0.49 & 0.49 & 0.01 & 0.58 & 29 & $10^{7}$ & 0.5 & 400 & 15.21 & 15.21 & 15.21 & 0.112 \\
CRB & 8 & 4 & 1536 & 768 & 256 & 0.56 & 0.56 & 0.01 & 0.65 & 29 & $10^{7}$ & 0.5 & 500 & 15.10 & 15.11 & 15.09 & 0.0918 \\
CRB & 8 & 4 & 1536 & 768 & 256 & 0.62 & 0.62 & 0.01 & 0.73 & 29 & $10^{7}$ & 0.5 & 600 & 15.06 & 15.06 & 15.08 & 0.0786 \\
CRB & 8 & 4 & 1536 & 768 & 256 & 0.68 & 0.68 & 0.01 & 0.79 & 29 & $10^{7}$ & 0.5 & 700 & 14.89 & 14.90 & 14.89 & 0.0688 \\
CRB & 8 & 4 & 1536 & 768 & 256 & 0.73 & 0.73 & 0.01 & 0.85 & 29 & $10^{7}$ & 0.5 & 800 & 14.74 & 14.75 & 14.74 & 0.061 \\
CRB & 8 & 4 & 1536 & 768 & 256 & 0.78 & 0.78 & 0.01 & 0.91 & 29 & $10^{7}$ & 0.5 & 900 & 14.58 & 14.58 & 14.57 & 0.055 \\
CRB & 8 & 4 & 1536 & 768 & 256 & 0.82 & 0.82 & 0.01 & 0.96 & 30 & $10^{7}$ & 0.5 & 1000 & 14.48 & 14.47 & 14.47 & 0.049 \\
CRB & 8 & 4 & 1536 & 768 & 256 & 0.89 & 0.89 & 0.01 & 1.05 & 30 & $10^{7}$ & 0.5 & 1200 & 14.22 & 14.22 & 14.20 & 0.041 \\
CRB & 8 & 4 & 1536 & 768 & 256 & 1.03 & 1.03 & 0.01 & 1.21 & 31 & $10^{7}$ & 0.5 & 1500 & 13.64 & 13.65 & 13.63 & 0.035 \\
CRB & 8 & 4 & 1536 & 768 & 256 & 1.22 & 1.22 & 0.02 & 1.43 & 31 & $10^{7}$ & 0.5 & 2000 & 12.92 & 12.92 & 12.92 & 0.0272 \\
CRB & 8 & 4 & 1536 & 768 & 256 & 1.50 & 1.50 & 0.02 & 1.77 & 33 & $10^{7}$ & 0.5 & 3000 & 12.04 & 12.05 & 12.04 & 0.0185 \\
CRB & 8 & 4 & 1536 & 768 & 256 & 1.75 & 1.75 & 0.03 & 2.06 & 33 & $10^{7}$ & 0.5 & 4000 & 11.64 & 11.64 & 11.65 & 0.0141 \\
CRB & 8 & 4 & 1536 & 768 & 256 & 1.99 & 1.99 & 0.03 & 2.34 & 33 & $10^{7}$ & 0.5 & 5000 & 11.43 & 11.44 & 11.43 & 0.0117 \\
CRB & 8 & 4 & 1536 & 768 & 256 & 2.20 & 2.20 & 0.03 & 2.58 & 33 & $10^{7}$ & 0.5 & 6000 & 11.48 & 11.48 & 11.47 & 0.00994 \\
CRB & 8 & 4 & 1536 & 768 & 256 & 2.41 & 2.41 & 0.03 & 2.83 & 33 & $10^{7}$ & 0.5 & 7000 & 11.66 & 11.67 & 11.67 & 0.00873 \\
CRB & 8 & 4 & 1536 & 768 & 256 & 2.61 & 2.61 & 0.04 & 3.07 & 33 & $10^{7}$ & 0.5 & 8000 & 11.92 & 11.92 & 11.91 & 0.00787 \\
CRB & 8 & 4 & 1536 & 768 & 256 & 2.83 & 2.83 & 0.04 & 3.32 & 32 & $10^{7}$ & 0.5 & 9000 & 12.32 & 12.32 & 12.32 & 0.00729 \\
CRB & 8 & 4 & 1536 & 768 & 256 & 3.06 & 3.06 & 0.04 & 3.59 & 32 & $10^{7}$ & 0.5 & 10000 & 12.82 & 12.83 & 12.82 & 0.00689 \\
\hline
& & & & & & & & & & & & & & & & &  \\[-6.5pt]
CRB & 8 & 4 & 1536 & 768 & 256 & 0.04 & 0.04 & 0.00 & 0.05 & 28 & $10^{7}$ & 1.0 & 10 & 15.73 & 15.73 & 15.73 & 1.42 \\
CRB & 8 & 4 & 1536 & 768 & 256 & 0.06 & 0.06 & 0.00 & 0.06 & 28 & $10^{7}$ & 1.0 & 20 & 15.77 & 15.77 & 15.78 & 0.56 \\
CRB & 8 & 4 & 1536 & 768 & 256 & 0.16 & 0.16 & 0.00 & 0.18 & 28 & $10^{7}$ & 1.0 & 30 & 15.77 & 15.77 & 15.78 & 1.97 \\
CRB & 8 & 4 & 1536 & 768 & 256 & 0.14 & 0.14 & 0.00 & 0.16 & 28 & $10^{7}$ & 1.0 & 40 & 15.66 & 15.66 & 15.69 & 0.853 \\
CRB & 8 & 4 & 1536 & 768 & 256 & 0.16 & 0.16 & 0.00 & 0.19 & 28 & $10^{7}$ & 1.0 & 50 & 15.67 & 15.67 & 15.66 & 0.77 \\
CRB & 8 & 4 & 1536 & 768 & 256 & 0.18 & 0.18 & 0.00 & 0.21 & 28 & $10^{7}$ & 1.0 & 60 & 15.68 & 15.69 & 15.69 & 0.679 \\
CRB & 8 & 4 & 1536 & 768 & 256 & 0.16 & 0.16 & 0.00 & 0.19 & 28 & $10^{7}$ & 1.0 & 70 & 15.79 & 15.79 & 15.82 & 0.391 \\
CRB & 8 & 4 & 1536 & 768 & 256 & 0.19 & 0.19 & 0.00 & 0.23 & 28 & $10^{7}$ & 1.0 & 80 & 15.74 & 15.74 & 15.70 & 0.425 \\
CRB & 8 & 4 & 1536 & 768 & 256 & 0.21 & 0.21 & 0.00 & 0.24 & 28 & $10^{7}$ & 1.0 & 90 & 15.76 & 15.76 & 15.74 & 0.391 \\
CRB & 8 & 4 & 1536 & 768 & 256 & 0.21 & 0.21 & 0.00 & 0.25 & 28 & $10^{7}$ & 1.0 & 100 & 15.76 & 15.76 & 15.76 & 0.338 \\
CRB & 8 & 4 & 1536 & 768 & 256 & 0.30 & 0.30 & 0.00 & 0.35 & 28 & $10^{7}$ & 1.0 & 200 & 15.64 & 15.65 & 15.66 & 0.168 \\
CRB & 8 & 4 & 1536 & 768 & 256 & 0.37 & 0.37 & 0.01 & 0.43 & 28 & $10^{7}$ & 1.0 & 300 & 15.58 & 15.57 & 15.60 & 0.112 \\
CRB & 8 & 4 & 1536 & 768 & 256 & 0.44 & 0.44 & 0.01 & 0.51 & 29 & $10^{7}$ & 1.0 & 400 & 15.45 & 15.44 & 15.47 & 0.0878 \\
CRB & 8 & 4 & 1536 & 768 & 256 & 0.50 & 0.50 & 0.01 & 0.59 & 29 & $10^{7}$ & 1.0 & 500 & 15.08 & 15.07 & 15.08 & 0.0745 \\
CRB & 8 & 4 & 1536 & 768 & 256 & 0.55 & 0.55 & 0.01 & 0.64 & 29 & $10^{7}$ & 1.0 & 600 & 14.93 & 14.91 & 14.93 & 0.0615 \\
CRB & 8 & 4 & 1536 & 768 & 256 & 0.61 & 0.61 & 0.01 & 0.71 & 29 & $10^{7}$ & 1.0 & 700 & 14.62 & 14.62 & 14.62 & 0.0552 \\
CRB & 8 & 4 & 1536 & 768 & 256 & 0.66 & 0.66 & 0.01 & 0.77 & 30 & $10^{7}$ & 1.0 & 800 & 14.27 & 14.27 & 14.28 & 0.0497 \\
CRB & 8 & 4 & 1536 & 768 & 256 & 0.70 & 0.70 & 0.01 & 0.82 & 30 & $10^{7}$ & 1.0 & 900 & 14.02 & 14.03 & 14.03 & 0.0443 \\
CRB & 8 & 4 & 1536 & 768 & 256 & 0.75 & 0.75 & 0.01 & 0.88 & 30 & $10^{7}$ & 1.0 & 1000 & 13.78 & 13.77 & 13.78 & 0.0413 \\
CRB & 8 & 4 & 1536 & 768 & 256 & 0.91 & 0.91 & 0.01 & 1.07 & 31 & $10^{7}$ & 1.0 & 1414 & 12.93 & 12.93 & 12.93 & 0.0306 \\
CRB & 8 & 4 & 1536 & 768 & 256 & 1.17 & 1.17 & 0.02 & 1.37 & 33 & $10^{7}$ & 1.0 & 2236 & 11.97 & 11.98 & 11.96 & 0.0201 \\
CRB & 8 & 4 & 1536 & 768 & 256 & 1.39 & 1.39 & 0.02 & 1.64 & 33 & $10^{7}$ & 1.0 & 3162 & 11.62 & 11.62 & 11.61 & 0.0143 \\
CRB & 8 & 4 & 1536 & 768 & 256 & 1.69 & 1.69 & 0.02 & 1.98 & 33 & $10^{7}$ & 1.0 & 4472 & 11.77 & 11.77 & 11.76 & 0.0105 \\
CRB & 8 & 4 & 1536 & 768 & 256 & 2.30 & 2.30 & 0.03 & 2.69 & 30 & $10^{7}$ & 1.0 & 7071 & 13.74 & 13.74 & 13.69 & 0.00777 \\
CRB & 8 & 4 & 1536 & 768 & 256 & 3.08 & 3.08 & 0.04 & 3.62 & 27 & $10^{7}$ & 1.0 & 10000 & 17.44 & 17.44 & 17.50 & 0.00701 \\
\hline
& & & & & & & & & & & & & & & & &  \\[-6.5pt]
CRB & 8 & 4 & 1536 & 768 & 256 & 0.05 & 0.05 & 0.00 & 0.06 & 28 & $10^{7}$ & 3.0 & 10 & 16.03 & 16.03 & 16.04 & 1.98 \\
CRB & 8 & 4 & 1536 & 768 & 256 & 0.07 & 0.07 & 0.00 & 0.09 & 28 & $10^{7}$ & 3.0 & 20 & 16.07 & 16.07 & 16.07 & 1.03 \\
CRB & 8 & 4 & 1536 & 768 & 256 & 0.09 & 0.09 & 0.00 & 0.11 & 28 & $10^{7}$ & 3.0 & 30 & 16.03 & 16.02 & 16.02 & 0.717 \\
CRB & 8 & 4 & 1536 & 768 & 256 & 0.10 & 0.10 & 0.00 & 0.12 & 28 & $10^{7}$ & 3.0 & 40 & 16.06 & 16.06 & 16.06 & 0.503 \\
CRB & 8 & 4 & 1536 & 768 & 256 & 0.13 & 0.13 & 0.00 & 0.15 & 28 & $10^{7}$ & 3.0 & 50 & 16.05 & 16.05 & 16.07 & 0.469 \\
CRB & 8 & 4 & 1536 & 768 & 256 & 0.13 & 0.13 & 0.00 & 0.16 & 28 & $10^{7}$ & 3.0 & 60 & 16.07 & 16.06 & 16.07 & 0.367 \\
CRB & 8 & 4 & 1536 & 768 & 256 & 0.15 & 0.15 & 0.00 & 0.18 & 28 & $10^{7}$ & 3.0 & 70 & 16.03 & 16.03 & 16.04 & 0.345 \\
CRB & 8 & 4 & 1536 & 768 & 256 & 0.16 & 0.16 & 0.00 & 0.19 & 28 & $10^{7}$ & 3.0 & 80 & 16.00 & 16.00 & 16.01 & 0.296 \\
CRB & 8 & 4 & 1536 & 768 & 256 & 0.17 & 0.17 & 0.00 & 0.20 & 28 & $10^{7}$ & 3.0 & 90 & 15.98 & 15.97 & 15.97 & 0.269 \\
CRB & 8 & 4 & 1536 & 768 & 256 & 0.18 & 0.18 & 0.00 & 0.21 & 28 & $10^{7}$ & 3.0 & 100 & 15.95 & 15.95 & 15.97 & 0.237 \\
CRB & 8 & 4 & 1536 & 768 & 256 & 0.26 & 0.26 & 0.00 & 0.30 & 28 & $10^{7}$ & 3.0 & 200 & 15.56 & 15.55 & 15.58 & 0.12 \\
CRB & 8 & 4 & 1536 & 768 & 256 & 0.31 & 0.31 & 0.00 & 0.37 & 29 & $10^{7}$ & 3.0 & 300 & 15.04 & 15.04 & 15.04 & 0.0807 \\
CRB & 8 & 4 & 1536 & 768 & 256 & 0.37 & 0.37 & 0.01 & 0.43 & 30 & $10^{7}$ & 3.0 & 400 & 14.47 & 14.47 & 14.47 & 0.0622 \\
CRB & 8 & 4 & 1536 & 768 & 256 & 0.43 & 0.43 & 0.01 & 0.51 & 30 & $10^{7}$ & 3.0 & 500 & 13.85 & 13.85 & 13.85 & 0.0555 \\
CRB & 8 & 4 & 1536 & 768 & 256 & 0.49 & 0.49 & 0.01 & 0.57 & 31 & $10^{7}$ & 3.0 & 600 & 13.41 & 13.41 & 13.42 & 0.0482 \\
CRB & 8 & 4 & 1536 & 768 & 256 & 0.53 & 0.53 & 0.01 & 0.62 & 31 & $10^{7}$ & 3.0 & 700 & 13.09 & 13.09 & 13.09 & 0.0421 \\
CRB & 8 & 4 & 1536 & 768 & 256 & 0.57 & 0.57 & 0.01 & 0.67 & 31 & $10^{7}$ & 3.0 & 800 & 12.84 & 12.84 & 12.85 & 0.0372 \\
CRB & 8 & 4 & 1536 & 768 & 256 & 0.60 & 0.60 & 0.01 & 0.71 & 32 & $10^{7}$ & 3.0 & 900 & 12.62 & 12.62 & 12.63 & 0.0333 \\
CRB & 8 & 4 & 1536 & 768 & 256 & 0.64 & 0.64 & 0.01 & 0.75 & 32 & $10^{7}$ & 3.0 & 1000 & 12.47 & 12.47 & 12.47 & 0.0301 \\
CRB & 8 & 4 & 1536 & 768 & 256 & 0.70 & 0.70 & 0.01 & 0.83 & 32 & $10^{7}$ & 3.0 & 1200 & 12.17 & 12.16 & 12.13 & 0.0254 \\
CRB & 8 & 4 & 1536 & 768 & 256 & 0.79 & 0.79 & 0.01 & 0.92 & 33 & $10^{7}$ & 3.0 & 1500 & 12.05 & 12.06 & 12.06 & 0.0204 \\
CRB & 8 & 4 & 1536 & 768 & 256 & 0.95 & 0.95 & 0.01 & 1.12 & 32 & $10^{7}$ & 3.0 & 2000 & 12.81 & 12.81 & 12.80 & 0.0167 \\
CRB & 8 & 4 & 1536 & 768 & 256 & 1.14 & 1.14 & 0.02 & 1.34 & 32 & $10^{7}$ & 3.0 & 3000 & 12.59 & 12.59 & 12.58 & 0.0106 \\
CRB & 8 & 4 & 1536 & 768 & 256 & 1.38 & 1.38 & 0.02 & 1.62 & 30 & $10^{7}$ & 3.0 & 4000 & 14.00 & 14.00 & 14.04 & 0.00878 \\
CRB & 8 & 4 & 1536 & 768 & 256 & 1.64 & 1.64 & 0.02 & 1.92 & 28 & $10^{7}$ & 3.0 & 5000 & 15.98 & 15.97 & 15.97 & 0.00788 \\
CRB & 8 & 4 & 1536 & 768 & 256 & 1.90 & 1.90 & 0.03 & 2.23 & 26 & $10^{7}$ & 3.0 & 6000 & 18.15 & 18.16 & 18.18 & 0.00739 \\
CRB & 8 & 4 & 1536 & 768 & 256 & 2.16 & 2.16 & 0.03 & 2.54 & 25 & $10^{7}$ & 3.0 & 7000 & 20.34 & 20.31 & 20.29 & 0.00705 \\
CRB & 8 & 4 & 1536 & 768 & 256 & 2.44 & 2.44 & 0.03 & 2.86 & 23 & $10^{7}$ & 3.0 & 8000 & 22.73 & 22.74 & 22.71 & 0.00683 \\
CRB & 8 & 4 & 1536 & 768 & 256 & 2.63 & 2.63 & 0.04 & 3.08 & 23 & $10^{7}$ & 3.0 & 9000 & 24.18 & 24.11 & 24.00 & 0.00629 \\
CRB & 8 & 4 & 1536 & 768 & 256 & 2.87 & 2.87 & 0.04 & 3.37 & 22 & $10^{7}$ & 3.0 & 10000 & 26.23 & 26.36 & 26.93 & 0.00609 \\
\hline
& & & & & & & & & & & & & & & & &  \\[-6.5pt]
CRB & 8 & 4 & 2048 & 1024 & 384 & 0.04 & 0.04 & 0.00 & 0.04 & 41 & $10^{7}$ & 5.0 & 10 & 16.13 & 16.13 & 16.13 & 2.13 \\
CRB & 8 & 4 & 2048 & 1024 & 384 & 0.06 & 0.06 & 0.00 & 0.06 & 41 & $10^{7}$ & 5.0 & 20 & 16.14 & 16.14 & 16.15 & 1.03 \\
CRB & 8 & 4 & 2048 & 1024 & 384 & 0.06 & 0.06 & 0.00 & 0.06 & 41 & $10^{7}$ & 5.0 & 30 & 16.13 & 16.12 & 16.14 & 0.553 \\
CRB & 8 & 4 & 2048 & 1024 & 384 & 0.08 & 0.08 & 0.00 & 0.08 & 41 & $10^{7}$ & 5.0 & 40 & 16.11 & 16.11 & 16.11 & 0.481 \\
CRB & 8 & 4 & 2048 & 1024 & 384 & 0.08 & 0.08 & 0.00 & 0.09 & 41 & $10^{7}$ & 5.0 & 50 & 16.09 & 16.09 & 16.08 & 0.368 \\
CRB & 8 & 4 & 2048 & 1024 & 384 & 0.09 & 0.09 & 0.00 & 0.10 & 41 & $10^{7}$ & 5.0 & 60 & 16.04 & 16.04 & 16.03 & 0.324 \\
CRB & 8 & 4 & 2048 & 1024 & 384 & 0.10 & 0.10 & 0.00 & 0.10 & 41 & $10^{7}$ & 5.0 & 70 & 16.02 & 16.02 & 16.02 & 0.264 \\
CRB & 8 & 4 & 2048 & 1024 & 384 & 0.11 & 0.11 & 0.00 & 0.11 & 41 & $10^{7}$ & 5.0 & 80 & 15.93 & 15.93 & 15.91 & 0.253 \\
CRB & 8 & 4 & 2048 & 1024 & 384 & 0.12 & 0.12 & 0.00 & 0.12 & 41 & $10^{7}$ & 5.0 & 90 & 15.98 & 15.98 & 15.97 & 0.217 \\
CRB & 8 & 4 & 2048 & 1024 & 384 & 0.12 & 0.12 & 0.00 & 0.13 & 41 & $10^{7}$ & 5.0 & 100 & 15.85 & 15.85 & 15.85 & 0.201 \\
CRB & 8 & 4 & 2048 & 1024 & 384 & 0.18 & 0.18 & 0.00 & 0.19 & 42 & $10^{7}$ & 5.0 & 200 & 15.09 & 15.09 & 15.10 & 0.108 \\
CRB & 8 & 4 & 2048 & 1024 & 384 & 0.23 & 0.23 & 0.01 & 0.23 & 44 & $10^{7}$ & 5.0 & 300 & 14.33 & 14.33 & 14.34 & 0.0751 \\
CRB & 8 & 4 & 2048 & 1024 & 384 & 0.26 & 0.26 & 0.01 & 0.27 & 45 & $10^{7}$ & 5.0 & 400 & 13.78 & 13.78 & 13.78 & 0.0552 \\
CRB & 8 & 4 & 2048 & 1024 & 384 & 0.31 & 0.31 & 0.01 & 0.32 & 46 & $10^{7}$ & 5.0 & 500 & 13.18 & 13.18 & 13.18 & 0.0492 \\
CRB & 8 & 4 & 2048 & 1024 & 384 & 0.34 & 0.34 & 0.01 & 0.35 & 46 & $10^{7}$ & 5.0 & 600 & 12.91 & 12.91 & 13.00 & 0.0411 \\
CRB & 8 & 4 & 2048 & 1024 & 384 & 0.36 & 0.36 & 0.01 & 0.38 & 47 & $10^{7}$ & 5.0 & 700 & 12.67 & 12.67 & 12.77 & 0.0354 \\
CRB & 8 & 4 & 2048 & 1024 & 384 & 0.40 & 0.40 & 0.01 & 0.42 & 46 & $10^{7}$ & 5.0 & 800 & 12.84 & 12.84 & 12.87 & 0.0332 \\
CRB & 8 & 4 & 2048 & 1024 & 384 & 0.43 & 0.43 & 0.01 & 0.45 & 46 & $10^{7}$ & 5.0 & 900 & 12.94 & 12.94 & 12.89 & 0.0303 \\
CRB & 8 & 4 & 2048 & 1024 & 384 & 0.46 & 0.46 & 0.01 & 0.48 & 46 & $10^{7}$ & 5.0 & 1000 & 13.07 & 13.07 & 13.04 & 0.028 \\
CRB & 8 & 4 & 2048 & 1024 & 384 & 0.48 & 0.48 & 0.01 & 0.50 & 48 & $10^{7}$ & 5.0 & 1200 & 12.20 & 12.20 & 12.21 & 0.0211 \\
CRB & 8 & 4 & 2048 & 1024 & 384 & 0.54 & 0.54 & 0.01 & 0.56 & 48 & $10^{7}$ & 5.0 & 1500 & 12.19 & 12.19 & 12.19 & 0.017 \\
CRB & 8 & 4 & 2048 & 1024 & 384 & 0.62 & 0.62 & 0.02 & 0.64 & 47 & $10^{7}$ & 5.0 & 2000 & 12.28 & 12.28 & 12.27 & 0.0128 \\
CRB & 8 & 4 & 2048 & 1024 & 384 & 0.80 & 0.80 & 0.02 & 0.83 & 44 & $10^{7}$ & 5.0 & 3000 & 13.98 & 13.98 & 13.96 & 0.00942 \\
CRB & 8 & 4 & 2048 & 1024 & 384 & 1.00 & 1.00 & 0.03 & 1.03 & 40 & $10^{7}$ & 5.0 & 4000 & 16.44 & 16.44 & 16.39 & 0.00818 \\
CRB & 8 & 4 & 2048 & 1024 & 384 & 1.20 & 1.20 & 0.03 & 1.24 & 37 & $10^{7}$ & 5.0 & 5000 & 19.14 & 19.13 & 19.14 & 0.0075 \\
CRB & 8 & 4 & 2048 & 1024 & 384 & 1.39 & 1.39 & 0.03 & 1.43 & 35 & $10^{7}$ & 5.0 & 6000 & 21.72 & 21.72 & 21.88 & 0.007 \\
CRB & 8 & 4 & 2048 & 1024 & 384 & 1.56 & 1.56 & 0.04 & 1.61 & 33 & $10^{7}$ & 5.0 & 7000 & 24.23 & 24.24 & 24.31 & 0.0065 \\
CRB & 8 & 4 & 2048 & 1024 & 384 & 1.76 & 1.76 & 0.04 & 1.82 & 31 & $10^{7}$ & 5.0 & 8000 & 27.10 & 27.08 & 27.10 & 0.00637 \\
CRB & 8 & 4 & 2048 & 1024 & 384 & 1.94 & 1.94 & 0.05 & 2.01 & 29 & $10^{7}$ & 5.0 & 9000 & 29.63 & 29.60 & 29.53 & 0.0061 \\
CRB & 8 & 4 & 2048 & 1024 & 384 & 2.15 & 2.15 & 0.05 & 2.22 & 28 & $10^{7}$ & 5.0 & 10000 & 32.59 & 32.54 & 32.42 & 0.00603 \\
\hline
& & & & & & & & & & & & & & & & &  \\[-6.5pt]
CRB & 8 & 4 & 2048 & 1024 & 384 & 0.26 & 0.26 & 0.01 & 0.27 & 29 & $10^{8}$ & 1.0 & 100 & 30.85 & 30.86 & 30.85 & 0.893 \\
CRB & 8 & 4 & 2048 & 1024 & 384 & 0.29 & 0.29 & 0.01 & 0.30 & 29 & $10^{8}$ & 1.0 & 200 & 30.68 & 30.68 & 30.78 & 0.281 \\
CRB & 8 & 4 & 2048 & 1024 & 384 & 0.36 & 0.36 & 0.01 & 0.37 & 29 & $10^{8}$ & 1.0 & 300 & 30.80 & 30.80 & 30.84 & 0.188 \\
CRB & 8 & 4 & 2048 & 1024 & 384 & 0.47 & 0.47 & 0.01 & 0.49 & 29 & $10^{8}$ & 1.0 & 400 & 30.81 & 30.80 & 30.87 & 0.183 \\
CRB & 8 & 4 & 2048 & 1024 & 384 & 0.51 & 0.51 & 0.01 & 0.53 & 29 & $10^{8}$ & 1.0 & 500 & 30.79 & 30.79 & 30.81 & 0.138 \\
CRB & 8 & 4 & 2048 & 1024 & 384 & 0.58 & 0.58 & 0.01 & 0.60 & 29 & $10^{8}$ & 1.0 & 600 & 30.72 & 30.73 & 30.74 & 0.124 \\
CRB & 8 & 4 & 2048 & 1024 & 384 & 0.63 & 0.63 & 0.02 & 0.65 & 29 & $10^{8}$ & 1.0 & 700 & 30.69 & 30.70 & 30.64 & 0.105 \\
CRB & 8 & 4 & 2048 & 1024 & 384 & 0.66 & 0.66 & 0.02 & 0.68 & 29 & $10^{8}$ & 1.0 & 800 & 30.66 & 30.65 & 30.71 & 0.0879 \\
CRB & 8 & 4 & 2048 & 1024 & 384 & 0.72 & 0.72 & 0.02 & 0.74 & 29 & $10^{8}$ & 1.0 & 900 & 30.55 & 30.54 & 30.49 & 0.0836 \\
CRB & 8 & 4 & 2048 & 1024 & 384 & 0.75 & 0.75 & 0.02 & 0.78 & 29 & $10^{8}$ & 1.0 & 1000 & 30.49 & 30.53 & 30.55 & 0.0743 \\
CRB & 8 & 4 & 2048 & 1024 & 384 & 0.80 & 0.80 & 0.02 & 0.83 & 29 & $10^{8}$ & 1.0 & 1200 & 30.45 & 30.45 & 30.50 & 0.0587 \\
CRB & 8 & 4 & 2048 & 1024 & 384 & 0.92 & 0.92 & 0.02 & 0.95 & 29 & $10^{8}$ & 1.0 & 1400 & 29.97 & 30.10 & 29.66 & 0.056 \\
CRB & 8 & 4 & 2048 & 1024 & 384 & 0.97 & 0.97 & 0.02 & 1.00 & 29 & $10^{8}$ & 1.0 & 1600 & 29.80 & 29.80 & 29.77 & 0.0481 \\
CRB & 8 & 4 & 2048 & 1024 & 384 & 1.03 & 1.03 & 0.03 & 1.06 & 29 & $10^{8}$ & 1.0 & 1800 & 29.54 & 29.53 & 29.52 & 0.0426 \\
CRB & 8 & 4 & 2048 & 1024 & 384 & 1.09 & 1.09 & 0.03 & 1.13 & 30 & $10^{8}$ & 1.0 & 2000 & 29.19 & 29.20 & 29.20 & 0.0392 \\
CRB & 8 & 4 & 2048 & 1024 & 384 & 1.17 & 1.17 & 0.03 & 1.21 & 30 & $10^{8}$ & 1.0 & 2200 & 28.73 & 28.74 & 28.67 & 0.0371 \\
CRB & 8 & 4 & 2048 & 1024 & 384 & 1.25 & 1.25 & 0.03 & 1.29 & 30 & $10^{8}$ & 1.0 & 2400 & 28.46 & 28.57 & 27.48 & 0.0356 \\
CRB & 8 & 4 & 2048 & 1024 & 384 & 1.28 & 1.28 & 0.03 & 1.32 & 30 & $10^{8}$ & 1.0 & 2600 & 28.16 & 28.16 & 28.11 & 0.0318 \\
CRB & 8 & 4 & 2048 & 1024 & 384 & 1.32 & 1.32 & 0.03 & 1.37 & 30 & $10^{8}$ & 1.0 & 2800 & 28.07 & 28.07 & 28.07 & 0.0292 \\
CRB & 8 & 4 & 2048 & 1024 & 384 & 1.37 & 1.37 & 0.03 & 1.42 & 30 & $10^{8}$ & 1.0 & 2900 & 27.82 & 27.83 & 27.87 & 0.0292 \\
CRB & 8 & 4 & 2048 & 1024 & 384 & 1.40 & 1.40 & 0.03 & 1.44 & 30 & $10^{8}$ & 1.0 & 3000 & 27.68 & 27.68 & 27.67 & 0.0284 \\
CRB & 8 & 4 & 2048 & 1024 & 384 & 1.44 & 1.44 & 0.04 & 1.49 & 31 & $10^{8}$ & 1.0 & 3200 & 27.40 & 27.42 & 27.48 & 0.0265 \\
CRB & 8 & 4 & 2048 & 1024 & 384 & 1.49 & 1.49 & 0.04 & 1.54 & 31 & $10^{8}$ & 1.0 & 3400 & 27.16 & 27.16 & 27.02 & 0.0252 \\
CRB & 8 & 4 & 2048 & 1024 & 384 & 1.54 & 1.54 & 0.04 & 1.60 & 31 & $10^{8}$ & 1.0 & 3600 & 26.99 & 27.00 & 27.01 & 0.0241 \\
CRB & 8 & 4 & 2048 & 1024 & 384 & 1.60 & 1.60 & 0.04 & 1.65 & 31 & $10^{8}$ & 1.0 & 3800 & 26.80 & 26.80 & 26.74 & 0.0232 \\
CRB & 8 & 4 & 2048 & 1024 & 384 & 1.64 & 1.64 & 0.04 & 1.69 & 31 & $10^{8}$ & 1.0 & 4000 & 26.60 & 26.63 & 26.58 & 0.022 \\
CRB & 8 & 4 & 2048 & 1024 & 384 & 1.85 & 1.85 & 0.05 & 1.91 & 32 & $10^{8}$ & 1.0 & 5000 & 26.05 & 26.07 & 26.18 & 0.0179 \\
CRB & 8 & 4 & 2048 & 1024 & 384 & 2.05 & 2.05 & 0.05 & 2.12 & 32 & $10^{8}$ & 1.0 & 6000 & 25.57 & 25.60 & 25.77 & 0.0153 \\
CRB & 8 & 4 & 2048 & 1024 & 384 & 2.24 & 2.24 & 0.06 & 2.32 & 32 & $10^{8}$ & 1.0 & 7000 & 25.19 & 25.19 & 25.19 & 0.0134 \\
CRB & 8 & 4 & 2048 & 1024 & 384 & 2.44 & 2.44 & 0.06 & 2.52 & 32 & $10^{8}$ & 1.0 & 8000 & 25.32 & 25.33 & 25.36 & 0.0121 \\
CRB & 8 & 4 & 2048 & 1024 & 384 & 2.60 & 2.60 & 0.07 & 2.69 & 32 & $10^{8}$ & 1.0 & 9000 & 25.49 & 25.49 & 25.61 & 0.011 \\
CRB & 8 & 4 & 2048 & 1024 & 384 & 2.77 & 2.77 & 0.07 & 2.86 & 32 & $10^{8}$ & 1.0 & 10000 & 25.69 & 25.68 & 25.64 & 0.0101 \\
\hline
& & & & & & & & & & & & & & & & &  \\[-6.5pt]
PRB & 8 & 4 & 1536 & 768 & 256 & 0.05 & 0.05 & 0.00 & 0.06 & 40 & $10^{6}$ & 1.0 & 10 & 8.25 & 8.25 & 8.25 & 2.15 \\
PRB & 8 & 4 & 1536 & 768 & 256 & 0.08 & 0.08 & 0.00 & 0.09 & 40 & $10^{6}$ & 1.0 & 20 & 8.31 & 8.31 & 8.31 & 1.15 \\
PRB & 8 & 4 & 1536 & 768 & 256 & 0.09 & 0.09 & 0.00 & 0.10 & 40 & $10^{6}$ & 1.0 & 30 & 8.30 & 8.31 & 8.31 & 0.644 \\
PRB & 8 & 4 & 1536 & 768 & 256 & 0.11 & 0.11 & 0.00 & 0.13 & 40 & $10^{6}$ & 1.0 & 40 & 8.28 & 8.28 & 8.28 & 0.524 \\
PRB & 8 & 4 & 1536 & 768 & 256 & 0.12 & 0.12 & 0.00 & 0.14 & 40 & $10^{6}$ & 1.0 & 50 & 8.29 & 8.29 & 8.29 & 0.417 \\
PRB & 8 & 4 & 1536 & 768 & 256 & 0.13 & 0.13 & 0.00 & 0.16 & 40 & $10^{6}$ & 1.0 & 60 & 8.26 & 8.26 & 8.26 & 0.358 \\
PRB & 8 & 4 & 1536 & 768 & 256 & 0.14 & 0.14 & 0.00 & 0.16 & 40 & $10^{6}$ & 1.0 & 70 & 8.25 & 8.25 & 8.25 & 0.294 \\
PRB & 8 & 4 & 1536 & 768 & 256 & 0.15 & 0.15 & 0.00 & 0.18 & 40 & $10^{6}$ & 1.0 & 80 & 8.19 & 8.19 & 8.19 & 0.259 \\
PRB & 8 & 4 & 1536 & 768 & 256 & 0.16 & 0.16 & 0.00 & 0.19 & 40 & $10^{6}$ & 1.0 & 90 & 8.21 & 8.21 & 8.21 & 0.237 \\
PRB & 8 & 4 & 1536 & 768 & 256 & 0.17 & 0.17 & 0.00 & 0.20 & 40 & $10^{6}$ & 1.0 & 100 & 8.17 & 8.16 & 8.17 & 0.216 \\
PRB & 8 & 4 & 1536 & 768 & 256 & 0.24 & 0.24 & 0.00 & 0.29 & 41 & $10^{6}$ & 1.0 & 200 & 7.86 & 7.87 & 7.87 & 0.11 \\
PRB & 8 & 4 & 1536 & 768 & 256 & 0.30 & 0.30 & 0.00 & 0.35 & 42 & $10^{6}$ & 1.0 & 300 & 7.37 & 7.36 & 7.36 & 0.0748 \\
PRB & 8 & 4 & 1536 & 768 & 256 & 0.36 & 0.36 & 0.01 & 0.42 & 43 & $10^{6}$ & 1.0 & 400 & 7.04 & 7.04 & 7.04 & 0.0595 \\
PRB & 8 & 4 & 1536 & 768 & 256 & 0.41 & 0.41 & 0.01 & 0.48 & 44 & $10^{6}$ & 1.0 & 500 & 6.74 & 6.74 & 6.74 & 0.0486 \\
PRB & 8 & 4 & 1536 & 768 & 256 & 0.45 & 0.45 & 0.01 & 0.53 & 45 & $10^{6}$ & 1.0 & 600 & 6.56 & 6.56 & 6.56 & 0.0414 \\
PRB & 8 & 4 & 1536 & 768 & 256 & 0.49 & 0.49 & 0.01 & 0.57 & 45 & $10^{6}$ & 1.0 & 700 & 6.43 & 6.43 & 6.44 & 0.0361 \\
PRB & 8 & 4 & 1536 & 768 & 256 & 0.53 & 0.53 & 0.01 & 0.62 & 45 & $10^{6}$ & 1.0 & 800 & 6.39 & 6.38 & 6.38 & 0.0322 \\
PRB & 8 & 4 & 1536 & 768 & 256 & 0.57 & 0.57 & 0.01 & 0.67 & 46 & $10^{6}$ & 1.0 & 900 & 6.35 & 6.35 & 6.35 & 0.0293 \\
PRB & 8 & 4 & 1536 & 768 & 256 & 0.60 & 0.60 & 0.01 & 0.70 & 46 & $10^{6}$ & 1.0 & 1000 & 6.32 & 6.31 & 6.32 & 0.0266 \\
PRB & 8 & 4 & 1536 & 768 & 256 & 0.67 & 0.67 & 0.01 & 0.78 & 46 & $10^{6}$ & 1.0 & 1200 & 6.31 & 6.30 & 6.30 & 0.0227 \\
PRB & 8 & 4 & 1536 & 768 & 256 & 0.80 & 0.80 & 0.01 & 0.94 & 43 & $10^{6}$ & 1.0 & 1500 & 7.07 & 7.06 & 7.06 & 0.0211 \\
PRB & 8 & 4 & 1536 & 768 & 256 & 0.95 & 0.95 & 0.01 & 1.11 & 42 & $10^{6}$ & 1.0 & 2000 & 7.34 & 7.32 & 7.33 & 0.0166 \\
PRB & 8 & 4 & 1536 & 768 & 256 & 1.13 & 1.13 & 0.02 & 1.32 & 44 & $10^{6}$ & 1.0 & 3000 & 6.67 & 6.65 & 6.64 & 0.0104 \\
PRB & 8 & 4 & 1536 & 768 & 256 & 1.37 & 1.37 & 0.02 & 1.61 & 41 & $10^{6}$ & 1.0 & 4000 & 7.59 & 7.57 & 7.55 & 0.00866 \\
PRB & 8 & 4 & 1536 & 768 & 256 & 1.68 & 1.68 & 0.02 & 1.97 & 39 & $10^{6}$ & 1.0 & 5000 & 8.50 & 8.45 & 8.49 & 0.0083 \\
PRB & 8 & 4 & 1536 & 768 & 256 & 1.99 & 1.99 & 0.03 & 2.34 & 36 & $10^{6}$ & 1.0 & 6000 & 10.05 & 9.99 & 10.08 & 0.00811 \\
PRB & 8 & 4 & 1536 & 768 & 256 & 2.28 & 2.28 & 0.03 & 2.68 & 34 & $10^{6}$ & 1.0 & 7000 & 11.10 & 11.03 & 11.05 & 0.00785 \\
PRB & 8 & 4 & 1536 & 768 & 256 & 2.56 & 2.56 & 0.04 & 3.01 & 33 & $10^{6}$ & 1.0 & 8000 & 11.64 & 11.54 & 11.60 & 0.00755 \\
PRB & 8 & 4 & 1536 & 768 & 256 & 2.83 & 2.83 & 0.04 & 3.32 & 33 & $10^{6}$ & 1.0 & 9000 & 11.68 & 11.55 & 11.60 & 0.00731 \\
PRB & 8 & 4 & 1536 & 768 & 256 & 3.10 & 3.10 & 0.04 & 3.63 & 32 & $10^{6}$ & 1.0 & 10000 & 12.35 & 12.16 & 12.25 & 0.00707 \\
\hline
& & & & & & & & & & & & & & & & &  \\[-6.5pt]
PRB & 8 & 4 & 1536 & 768 & 256 & 0.08 & 0.08 & 0.00 & 0.09 & 29 & $10^{7}$ & 0.5 & 10 & 15.27 & 15.27 & 15.27 & 4.76 \\
PRB & 8 & 4 & 1536 & 768 & 256 & 0.12 & 0.12 & 0.00 & 0.14 & 29 & $10^{7}$ & 0.5 & 20 & 15.29 & 15.30 & 15.30 & 2.57 \\
PRB & 8 & 4 & 1536 & 768 & 256 & 0.14 & 0.14 & 0.00 & 0.16 & 29 & $10^{7}$ & 0.5 & 30 & 15.27 & 15.27 & 15.27 & 1.56 \\
PRB & 8 & 4 & 1536 & 768 & 256 & 0.15 & 0.15 & 0.00 & 0.18 & 29 & $10^{7}$ & 0.5 & 40 & 15.24 & 15.25 & 15.25 & 1.07 \\
PRB & 8 & 4 & 1536 & 768 & 256 & 0.18 & 0.18 & 0.00 & 0.21 & 29 & $10^{7}$ & 0.5 & 50 & 15.28 & 15.29 & 15.28 & 0.923 \\
PRB & 8 & 4 & 1536 & 768 & 256 & 0.20 & 0.20 & 0.00 & 0.23 & 29 & $10^{7}$ & 0.5 & 60 & 15.26 & 15.26 & 15.27 & 0.82 \\
PRB & 8 & 4 & 1536 & 768 & 256 & 0.22 & 0.22 & 0.00 & 0.25 & 29 & $10^{7}$ & 0.5 & 70 & 15.29 & 15.29 & 15.30 & 0.708 \\
PRB & 8 & 4 & 1536 & 768 & 256 & 0.23 & 0.23 & 0.00 & 0.27 & 29 & $10^{7}$ & 0.5 & 80 & 15.23 & 15.23 & 15.23 & 0.607 \\
PRB & 8 & 4 & 1536 & 768 & 256 & 0.25 & 0.25 & 0.00 & 0.30 & 29 & $10^{7}$ & 0.5 & 90 & 15.26 & 15.27 & 15.26 & 0.588 \\
PRB & 8 & 4 & 1536 & 768 & 256 & 0.26 & 0.26 & 0.00 & 0.31 & 29 & $10^{7}$ & 0.5 & 100 & 15.30 & 15.31 & 15.30 & 0.511 \\
PRB & 8 & 4 & 1536 & 768 & 256 & 0.37 & 0.37 & 0.01 & 0.43 & 29 & $10^{7}$ & 0.5 & 200 & 15.20 & 15.21 & 15.21 & 0.249 \\
PRB & 8 & 4 & 1536 & 768 & 256 & 0.44 & 0.44 & 0.01 & 0.52 & 29 & $10^{7}$ & 0.5 & 300 & 15.21 & 15.21 & 15.22 & 0.159 \\
PRB & 8 & 4 & 1536 & 768 & 256 & 0.51 & 0.51 & 0.01 & 0.60 & 29 & $10^{7}$ & 0.5 & 400 & 15.12 & 15.12 & 15.13 & 0.121 \\
PRB & 8 & 4 & 1536 & 768 & 256 & 0.57 & 0.57 & 0.01 & 0.67 & 29 & $10^{7}$ & 0.5 & 500 & 15.09 & 15.10 & 15.09 & 0.0974 \\
PRB & 8 & 4 & 1536 & 768 & 256 & 0.62 & 0.62 & 0.01 & 0.73 & 29 & $10^{7}$ & 0.5 & 600 & 15.02 & 15.02 & 15.03 & 0.0799 \\
PRB & 8 & 4 & 1536 & 768 & 256 & 0.68 & 0.68 & 0.01 & 0.80 & 29 & $10^{7}$ & 0.5 & 700 & 14.91 & 14.91 & 14.92 & 0.0695 \\
PRB & 8 & 4 & 1536 & 768 & 256 & 0.72 & 0.72 & 0.01 & 0.85 & 29 & $10^{7}$ & 0.5 & 800 & 14.82 & 14.81 & 14.81 & 0.06 \\
PRB & 8 & 4 & 1536 & 768 & 256 & 0.77 & 0.77 & 0.01 & 0.90 & 29 & $10^{7}$ & 0.5 & 900 & 14.67 & 14.67 & 14.67 & 0.0537 \\
PRB & 8 & 4 & 1536 & 768 & 256 & 0.81 & 0.81 & 0.01 & 0.95 & 29 & $10^{7}$ & 0.5 & 1000 & 14.59 & 14.60 & 14.60 & 0.0487 \\
PRB & 8 & 4 & 1536 & 768 & 256 & 0.90 & 0.90 & 0.01 & 1.06 & 30 & $10^{7}$ & 0.5 & 1200 & 14.31 & 14.31 & 14.31 & 0.0415 \\
PRB & 8 & 4 & 1536 & 768 & 256 & 1.02 & 1.02 & 0.01 & 1.19 & 30 & $10^{7}$ & 0.5 & 1500 & 13.85 & 13.85 & 13.86 & 0.0338 \\
PRB & 8 & 4 & 1536 & 768 & 256 & 1.19 & 1.19 & 0.02 & 1.40 & 31 & $10^{7}$ & 0.5 & 2000 & 13.20 & 13.20 & 13.21 & 0.0262 \\
PRB & 8 & 4 & 1536 & 768 & 256 & 1.49 & 1.49 & 0.02 & 1.75 & 32 & $10^{7}$ & 0.5 & 3000 & 12.24 & 12.23 & 12.24 & 0.0183 \\
PRB & 8 & 4 & 1536 & 768 & 256 & 1.74 & 1.74 & 0.03 & 2.05 & 33 & $10^{7}$ & 0.5 & 4000 & 11.77 & 11.77 & 11.78 & 0.014 \\
PRB & 8 & 4 & 1536 & 768 & 256 & 1.96 & 1.96 & 0.03 & 2.30 & 33 & $10^{7}$ & 0.5 & 5000 & 11.68 & 11.67 & 11.67 & 0.0114 \\
PRB & 8 & 4 & 1536 & 768 & 256 & 2.19 & 2.19 & 0.03 & 2.57 & 33 & $10^{7}$ & 0.5 & 6000 & 11.69 & 11.68 & 11.68 & 0.00981 \\
PRB & 8 & 4 & 1536 & 768 & 256 & 2.42 & 2.42 & 0.03 & 2.84 & 33 & $10^{7}$ & 0.5 & 7000 & 11.82 & 11.80 & 11.79 & 0.00878 \\
PRB & 8 & 4 & 1536 & 768 & 256 & 2.66 & 2.66 & 0.04 & 3.12 & 32 & $10^{7}$ & 0.5 & 8000 & 12.10 & 12.08 & 12.09 & 0.00812 \\
PRB & 8 & 4 & 1536 & 768 & 256 & 2.93 & 2.93 & 0.04 & 3.44 & 32 & $10^{7}$ & 0.5 & 9000 & 12.59 & 12.57 & 12.58 & 0.00781 \\
PRB & 8 & 4 & 1536 & 768 & 256 & 3.17 & 3.17 & 0.05 & 3.73 & 31 & $10^{7}$ & 0.5 & 10000 & 13.16 & 13.14 & 13.16 & 0.00743 \\
\hline
& & & & & & & & & & & & & & & & &  \\[-6.5pt]
PRB & 8 & 4 & 1536 & 768 & 256 & 0.07 & 0.07 & 0.00 & 0.09 & 28 & $10^{7}$ & 1.0 & 10 & 15.74 & 15.74 & 15.74 & 3.95 \\
PRB & 8 & 4 & 1536 & 768 & 256 & 0.10 & 0.10 & 0.00 & 0.11 & 28 & $10^{7}$ & 1.0 & 20 & 15.73 & 15.74 & 15.74 & 1.73 \\
PRB & 8 & 4 & 1536 & 768 & 256 & 0.16 & 0.16 & 0.00 & 0.19 & 28 & $10^{7}$ & 1.0 & 50 & 15.73 & 15.73 & 15.73 & 0.77 \\
PRB & 8 & 4 & 1536 & 768 & 256 & 0.23 & 0.23 & 0.00 & 0.27 & 28 & $10^{7}$ & 1.0 & 100 & 15.74 & 15.75 & 15.75 & 0.395 \\
PRB & 8 & 4 & 1536 & 768 & 256 & 0.31 & 0.31 & 0.00 & 0.37 & 28 & $10^{7}$ & 1.0 & 200 & 15.67 & 15.67 & 15.68 & 0.182 \\
PRB & 8 & 4 & 1536 & 768 & 256 & 0.39 & 0.39 & 0.01 & 0.45 & 29 & $10^{7}$ & 1.0 & 300 & 15.54 & 15.55 & 15.55 & 0.123 \\
PRB & 8 & 4 & 1536 & 768 & 256 & 0.45 & 0.45 & 0.01 & 0.52 & 29 & $10^{7}$ & 1.0 & 400 & 15.39 & 15.39 & 15.39 & 0.0913 \\
PRB & 8 & 4 & 1536 & 768 & 256 & 0.50 & 0.50 & 0.01 & 0.59 & 29 & $10^{7}$ & 1.0 & 500 & 15.15 & 15.15 & 15.14 & 0.0746 \\
PRB & 8 & 4 & 1536 & 768 & 256 & 0.55 & 0.55 & 0.01 & 0.65 & 29 & $10^{7}$ & 1.0 & 600 & 14.97 & 14.97 & 14.97 & 0.0625 \\
PRB & 8 & 4 & 1536 & 768 & 256 & 0.60 & 0.60 & 0.01 & 0.71 & 29 & $10^{7}$ & 1.0 & 700 & 14.71 & 14.71 & 14.71 & 0.0548 \\
PRB & 8 & 4 & 1536 & 768 & 256 & 0.65 & 0.65 & 0.01 & 0.76 & 30 & $10^{7}$ & 1.0 & 800 & 14.51 & 14.51 & 14.50 & 0.0488 \\
PRB & 8 & 4 & 1536 & 768 & 256 & 0.70 & 0.70 & 0.01 & 0.82 & 30 & $10^{7}$ & 1.0 & 900 & 14.21 & 14.21 & 14.21 & 0.0442 \\
PRB & 8 & 4 & 1536 & 768 & 256 & 0.74 & 0.74 & 0.01 & 0.87 & 30 & $10^{7}$ & 1.0 & 1000 & 13.99 & 13.99 & 14.00 & 0.0404 \\
PRB & 8 & 4 & 1536 & 768 & 256 & 0.90 & 0.90 & 0.01 & 1.06 & 31 & $10^{7}$ & 1.0 & 1414 & 13.17 & 13.17 & 13.18 & 0.03 \\
PRB & 8 & 4 & 1536 & 768 & 256 & 1.16 & 1.16 & 0.02 & 1.36 & 32 & $10^{7}$ & 1.0 & 2236 & 12.18 & 12.18 & 12.20 & 0.0197 \\
PRB & 8 & 4 & 1536 & 768 & 256 & 1.40 & 1.40 & 0.02 & 1.64 & 33 & $10^{7}$ & 1.0 & 3162 & 11.86 & 11.85 & 11.87 & 0.0144 \\
PRB & 8 & 4 & 1536 & 768 & 256 & 1.71 & 1.71 & 0.02 & 2.01 & 33 & $10^{7}$ & 1.0 & 4472 & 11.91 & 11.90 & 11.91 & 0.0108 \\
PRB & 8 & 4 & 1536 & 768 & 256 & 2.36 & 2.36 & 0.03 & 2.77 & 31 & $10^{7}$ & 1.0 & 7071 & 13.29 & 13.25 & 13.30 & 0.0082 \\
PRB & 8 & 4 & 1536 & 768 & 256 & 3.16 & 3.16 & 0.05 & 3.71 & 27 & $10^{7}$ & 1.0 & 10000 & 17.11 & 17.02 & 17.04 & 0.00735 \\
\hline
& & & & & & & & & & & & & & & & &  \\[-6.5pt]
PRB & 8 & 4 & 1536 & 768 & 256 & 0.05 & 0.05 & 0.00 & 0.06 & 28 & $10^{7}$ & 3.0 & 10 & 16.11 & 16.11 & 16.12 & 2.14 \\
PRB & 8 & 4 & 1536 & 768 & 256 & 0.08 & 0.08 & 0.00 & 0.10 & 28 & $10^{7}$ & 3.0 & 20 & 16.12 & 16.12 & 16.13 & 1.23 \\
PRB & 8 & 4 & 1536 & 768 & 256 & 0.10 & 0.10 & 0.00 & 0.12 & 28 & $10^{7}$ & 3.0 & 30 & 16.12 & 16.11 & 16.12 & 0.829 \\
PRB & 8 & 4 & 1536 & 768 & 256 & 0.12 & 0.12 & 0.00 & 0.14 & 28 & $10^{7}$ & 3.0 & 40 & 16.10 & 16.10 & 16.11 & 0.636 \\
PRB & 8 & 4 & 1536 & 768 & 256 & 0.13 & 0.13 & 0.00 & 0.15 & 28 & $10^{7}$ & 3.0 & 50 & 16.05 & 16.05 & 16.06 & 0.488 \\
PRB & 8 & 4 & 1536 & 768 & 256 & 0.15 & 0.15 & 0.00 & 0.17 & 28 & $10^{7}$ & 3.0 & 60 & 16.02 & 16.02 & 16.02 & 0.452 \\
PRB & 8 & 4 & 1536 & 768 & 256 & 0.15 & 0.15 & 0.00 & 0.18 & 28 & $10^{7}$ & 3.0 & 70 & 16.05 & 16.05 & 16.05 & 0.347 \\
PRB & 8 & 4 & 1536 & 768 & 256 & 0.16 & 0.16 & 0.00 & 0.19 & 28 & $10^{7}$ & 3.0 & 80 & 16.03 & 16.03 & 16.04 & 0.308 \\
PRB & 8 & 4 & 1536 & 768 & 256 & 0.17 & 0.17 & 0.00 & 0.20 & 28 & $10^{7}$ & 3.0 & 90 & 16.00 & 16.00 & 16.01 & 0.275 \\
PRB & 8 & 4 & 1536 & 768 & 256 & 0.18 & 0.18 & 0.00 & 0.22 & 28 & $10^{7}$ & 3.0 & 100 & 15.98 & 15.98 & 15.99 & 0.25 \\
PRB & 8 & 4 & 1536 & 768 & 256 & 0.26 & 0.26 & 0.00 & 0.31 & 28 & $10^{7}$ & 3.0 & 200 & 15.55 & 15.55 & 15.56 & 0.126 \\
PRB & 8 & 4 & 1536 & 768 & 256 & 0.33 & 0.33 & 0.00 & 0.39 & 29 & $10^{7}$ & 3.0 & 300 & 15.02 & 15.01 & 15.02 & 0.0897 \\
PRB & 8 & 4 & 1536 & 768 & 256 & 0.39 & 0.39 & 0.01 & 0.46 & 30 & $10^{7}$ & 3.0 & 400 & 14.43 & 14.43 & 14.44 & 0.0705 \\
PRB & 8 & 4 & 1536 & 768 & 256 & 0.45 & 0.45 & 0.01 & 0.52 & 30 & $10^{7}$ & 3.0 & 500 & 13.97 & 13.97 & 13.98 & 0.0585 \\
PRB & 8 & 4 & 1536 & 768 & 256 & 0.49 & 0.49 & 0.01 & 0.58 & 31 & $10^{7}$ & 3.0 & 600 & 13.61 & 13.60 & 13.62 & 0.0497 \\
PRB & 8 & 4 & 1536 & 768 & 256 & 0.54 & 0.54 & 0.01 & 0.63 & 31 & $10^{7}$ & 3.0 & 700 & 13.28 & 13.28 & 13.28 & 0.0436 \\
PRB & 8 & 4 & 1536 & 768 & 256 & 0.58 & 0.58 & 0.01 & 0.68 & 31 & $10^{7}$ & 3.0 & 800 & 13.01 & 13.01 & 13.02 & 0.0384 \\
PRB & 8 & 4 & 1536 & 768 & 256 & 0.62 & 0.62 & 0.01 & 0.72 & 32 & $10^{7}$ & 3.0 & 900 & 12.78 & 12.78 & 12.80 & 0.0345 \\
PRB & 8 & 4 & 1536 & 768 & 256 & 0.65 & 0.65 & 0.01 & 0.76 & 32 & $10^{7}$ & 3.0 & 1000 & 12.65 & 12.64 & 12.65 & 0.0313 \\
PRB & 8 & 4 & 1536 & 768 & 256 & 0.72 & 0.72 & 0.01 & 0.85 & 32 & $10^{7}$ & 3.0 & 1200 & 12.39 & 12.39 & 12.42 & 0.0266 \\
PRB & 8 & 4 & 1536 & 768 & 256 & 0.81 & 0.81 & 0.01 & 0.95 & 32 & $10^{7}$ & 3.0 & 1500 & 12.26 & 12.25 & 12.27 & 0.0217 \\
PRB & 8 & 4 & 1536 & 768 & 256 & 0.95 & 0.95 & 0.01 & 1.12 & 32 & $10^{7}$ & 3.0 & 2000 & 12.20 & 12.19 & 12.20 & 0.0167 \\
PRB & 8 & 4 & 1536 & 768 & 256 & 1.20 & 1.20 & 0.02 & 1.41 & 33 & $10^{7}$ & 3.0 & 3000 & 12.09 & 12.08 & 12.08 & 0.0119 \\
PRB & 8 & 4 & 1536 & 768 & 256 & 1.44 & 1.44 & 0.02 & 1.69 & 31 & $10^{7}$ & 3.0 & 4000 & 13.18 & 13.14 & 13.27 & 0.00954 \\
PRB & 8 & 4 & 1536 & 768 & 256 & 1.71 & 1.71 & 0.02 & 2.01 & 29 & $10^{7}$ & 3.0 & 5000 & 14.92 & 14.85 & 14.90 & 0.00865 \\
PRB & 8 & 4 & 1536 & 768 & 256 & 1.99 & 1.99 & 0.03 & 2.33 & 27 & $10^{7}$ & 3.0 & 6000 & 17.68 & 17.57 & 17.60 & 0.00808 \\
PRB & 8 & 4 & 1536 & 768 & 256 & 2.30 & 2.30 & 0.03 & 2.69 & 25 & $10^{7}$ & 3.0 & 7000 & 20.20 & 20.02 & 20.23 & 0.00793 \\
PRB & 8 & 4 & 1536 & 768 & 256 & 2.58 & 2.58 & 0.04 & 3.03 & 23 & $10^{7}$ & 3.0 & 8000 & 22.88 & 22.66 & 22.89 & 0.00768 \\
PRB & 8 & 4 & 1536 & 768 & 256 & 2.86 & 2.86 & 0.04 & 3.35 & 22 & $10^{7}$ & 3.0 & 9000 & 25.00 & 24.70 & 25.00 & 0.00743 \\
PRB & 8 & 4 & 1536 & 768 & 256 & 3.13 & 3.13 & 0.04 & 3.67 & 21 & $10^{7}$ & 3.0 & 10000 & 26.99 & 26.59 & 26.96 & 0.0072 \\
\hline
& & & & & & & & & & & & & & & & &  \\[-6.5pt]
PRB & 8 & 4 & 1536 & 768 & 256 & 0.05 & 0.05 & 0.00 & 0.06 & 28 & $10^{7}$ & 5.0 & 10 & 16.20 & 16.19 & 16.20 & 2.16 \\
PRB & 8 & 4 & 1536 & 768 & 256 & 0.08 & 0.08 & 0.00 & 0.09 & 28 & $10^{7}$ & 5.0 & 20 & 16.13 & 16.12 & 16.13 & 1.2 \\
PRB & 8 & 4 & 1536 & 768 & 256 & 0.10 & 0.10 & 0.00 & 0.11 & 28 & $10^{7}$ & 5.0 & 30 & 16.11 & 16.11 & 16.11 & 0.778 \\
PRB & 8 & 4 & 1536 & 768 & 256 & 0.11 & 0.11 & 0.00 & 0.13 & 28 & $10^{7}$ & 5.0 & 40 & 16.10 & 16.10 & 16.10 & 0.593 \\
PRB & 8 & 4 & 1536 & 768 & 256 & 0.12 & 0.12 & 0.00 & 0.14 & 28 & $10^{7}$ & 5.0 & 50 & 16.10 & 16.10 & 16.11 & 0.42 \\
PRB & 8 & 4 & 1536 & 768 & 256 & 0.13 & 0.13 & 0.00 & 0.16 & 28 & $10^{7}$ & 5.0 & 60 & 16.08 & 16.08 & 16.08 & 0.358 \\
PRB & 8 & 4 & 1536 & 768 & 256 & 0.15 & 0.15 & 0.00 & 0.18 & 28 & $10^{7}$ & 5.0 & 70 & 16.01 & 16.01 & 16.02 & 0.341 \\
PRB & 8 & 4 & 1536 & 768 & 256 & 0.15 & 0.15 & 0.00 & 0.18 & 28 & $10^{7}$ & 5.0 & 80 & 16.03 & 16.02 & 16.03 & 0.268 \\
PRB & 8 & 4 & 1536 & 768 & 256 & 0.16 & 0.16 & 0.00 & 0.19 & 28 & $10^{7}$ & 5.0 & 90 & 15.94 & 15.94 & 15.95 & 0.239 \\
PRB & 8 & 4 & 1536 & 768 & 256 & 0.17 & 0.17 & 0.00 & 0.20 & 28 & $10^{7}$ & 5.0 & 100 & 15.91 & 15.91 & 15.92 & 0.218 \\
PRB & 8 & 4 & 1536 & 768 & 256 & 0.25 & 0.25 & 0.00 & 0.30 & 29 & $10^{7}$ & 5.0 & 200 & 15.18 & 15.17 & 15.18 & 0.118 \\
PRB & 8 & 4 & 1536 & 768 & 256 & 0.32 & 0.32 & 0.00 & 0.38 & 30 & $10^{7}$ & 5.0 & 300 & 14.37 & 14.37 & 14.38 & 0.0838 \\
PRB & 8 & 4 & 1536 & 768 & 256 & 0.37 & 0.37 & 0.01 & 0.44 & 30 & $10^{7}$ & 5.0 & 400 & 13.78 & 13.78 & 13.79 & 0.0648 \\
PRB & 8 & 4 & 1536 & 768 & 256 & 0.42 & 0.42 & 0.01 & 0.50 & 31 & $10^{7}$ & 5.0 & 500 & 13.30 & 13.30 & 13.30 & 0.0528 \\
PRB & 8 & 4 & 1536 & 768 & 256 & 0.47 & 0.47 & 0.01 & 0.55 & 31 & $10^{7}$ & 5.0 & 600 & 12.99 & 12.99 & 13.00 & 0.0446 \\
PRB & 8 & 4 & 1536 & 768 & 256 & 0.51 & 0.51 & 0.01 & 0.60 & 32 & $10^{7}$ & 5.0 & 700 & 12.78 & 12.78 & 12.80 & 0.0387 \\
PRB & 8 & 4 & 1536 & 768 & 256 & 0.55 & 0.55 & 0.01 & 0.64 & 32 & $10^{7}$ & 5.0 & 800 & 12.59 & 12.59 & 12.61 & 0.0343 \\
PRB & 8 & 4 & 1536 & 768 & 256 & 0.58 & 0.58 & 0.01 & 0.68 & 32 & $10^{7}$ & 5.0 & 900 & 12.53 & 12.52 & 12.54 & 0.0307 \\
PRB & 8 & 4 & 1536 & 768 & 256 & 0.61 & 0.61 & 0.01 & 0.72 & 32 & $10^{7}$ & 5.0 & 1000 & 12.43 & 12.43 & 12.44 & 0.0278 \\
PRB & 8 & 4 & 1536 & 768 & 256 & 0.68 & 0.68 & 0.01 & 0.79 & 32 & $10^{7}$ & 5.0 & 1200 & 12.34 & 12.34 & 12.34 & 0.0234 \\
PRB & 8 & 4 & 1536 & 768 & 256 & 0.79 & 0.79 & 0.01 & 0.92 & 32 & $10^{7}$ & 5.0 & 1500 & 12.49 & 12.48 & 12.46 & 0.0203 \\
PRB & 8 & 4 & 1536 & 768 & 256 & 0.91 & 0.91 & 0.01 & 1.06 & 33 & $10^{7}$ & 5.0 & 2000 & 12.01 & 11.99 & 11.99 & 0.0151 \\
PRB & 8 & 4 & 1536 & 768 & 256 & 1.14 & 1.14 & 0.02 & 1.34 & 31 & $10^{7}$ & 5.0 & 3000 & 13.48 & 13.45 & 13.47 & 0.0107 \\
PRB & 8 & 4 & 1536 & 768 & 256 & 1.41 & 1.41 & 0.02 & 1.65 & 28 & $10^{7}$ & 5.0 & 4000 & 15.62 & 15.55 & 15.60 & 0.00912 \\
PRB & 8 & 4 & 1536 & 768 & 256 & 1.71 & 1.71 & 0.02 & 2.00 & 26 & $10^{7}$ & 5.0 & 5000 & 18.47 & 18.32 & 18.55 & 0.00858 \\
PRB & 8 & 4 & 1536 & 768 & 256 & 2.00 & 2.00 & 0.03 & 2.35 & 24 & $10^{7}$ & 5.0 & 6000 & 21.72 & 21.52 & 21.94 & 0.00822 \\
PRB & 8 & 4 & 1536 & 768 & 256 & 2.29 & 2.29 & 0.03 & 2.69 & 22 & $10^{7}$ & 5.0 & 7000 & 24.50 & 24.14 & 24.41 & 0.00791 \\
PRB & 8 & 4 & 1536 & 768 & 256 & 2.57 & 2.57 & 0.04 & 3.01 & 21 & $10^{7}$ & 5.0 & 8000 & 26.90 & 26.39 & 26.97 & 0.00759 \\
PRB & 8 & 4 & 1536 & 768 & 256 & 2.84 & 2.84 & 0.04 & 3.33 & 21 & $10^{7}$ & 5.0 & 9000 & 28.72 & 27.98 & 28.95 & 0.00733 \\
PRB & 8 & 4 & 1536 & 768 & 256 & 3.10 & 3.10 & 0.04 & 3.64 & 20 & $10^{7}$ & 5.0 & 10000 & 30.42 & 29.42 & 30.49 & 0.0071 \\
\hline
& & & & & & & & & & & & & & & & &  \\[-6.5pt]
PRB & 8 & 4 & 2048 & 1024 & 384 & 0.04 & 0.04 & 0.00 & 0.04 & 30 & $10^{8}$ & 1.0 & 10 & 30.73 & 30.75 & 30.76 & 2.17 \\
PRB & 8 & 4 & 2048 & 1024 & 384 & 0.14 & 0.14 & 0.00 & 0.15 & 30 & $10^{8}$ & 1.0 & 20 & 30.85 & 30.86 & 30.97 & 6.35 \\
PRB & 8 & 4 & 2048 & 1024 & 384 & 0.13 & 0.13 & 0.00 & 0.14 & 30 & $10^{8}$ & 1.0 & 30 & 30.78 & 30.78 & 30.86 & 2.55 \\
PRB & 8 & 4 & 2048 & 1024 & 384 & 0.16 & 0.16 & 0.00 & 0.17 & 30 & $10^{8}$ & 1.0 & 40 & 30.98 & 30.98 & 31.04 & 2.16 \\
PRB & 8 & 4 & 2048 & 1024 & 384 & 0.19 & 0.19 & 0.00 & 0.20 & 30 & $10^{8}$ & 1.0 & 50 & 30.92 & 30.92 & 30.90 & 1.89 \\
PRB & 8 & 4 & 2048 & 1024 & 384 & 0.18 & 0.18 & 0.00 & 0.19 & 30 & $10^{8}$ & 1.0 & 60 & 30.76 & 30.78 & 30.83 & 1.19 \\
PRB & 8 & 4 & 2048 & 1024 & 384 & 0.21 & 0.21 & 0.00 & 0.22 & 30 & $10^{8}$ & 1.0 & 70 & 30.87 & 30.87 & 30.86 & 1.22 \\
PRB & 8 & 4 & 2048 & 1024 & 384 & 0.23 & 0.23 & 0.00 & 0.24 & 30 & $10^{8}$ & 1.0 & 80 & 30.72 & 30.72 & 30.77 & 1.11 \\
PRB & 8 & 4 & 2048 & 1024 & 384 & 0.25 & 0.25 & 0.00 & 0.27 & 30 & $10^{8}$ & 1.0 & 90 & 30.69 & 30.70 & 30.73 & 1.05 \\
PRB & 8 & 4 & 2048 & 1024 & 384 & 0.21 & 0.21 & 0.00 & 0.22 & 30 & $10^{8}$ & 1.0 & 100 & 30.93 & 30.92 & 30.89 & 0.562 \\
PRB & 8 & 4 & 2048 & 1024 & 384 & 0.32 & 0.32 & 0.00 & 0.33 & 30 & $10^{8}$ & 1.0 & 200 & 30.85 & 30.85 & 30.83 & 0.329 \\
PRB & 8 & 4 & 2048 & 1024 & 384 & 0.42 & 0.42 & 0.00 & 0.44 & 30 & $10^{8}$ & 1.0 & 300 & 30.81 & 30.81 & 30.84 & 0.254 \\
PRB & 8 & 4 & 2048 & 1024 & 384 & 0.47 & 0.47 & 0.00 & 0.49 & 30 & $10^{8}$ & 1.0 & 400 & 30.69 & 30.68 & 30.61 & 0.184 \\
PRB & 8 & 4 & 2048 & 1024 & 384 & 0.53 & 0.53 & 0.00 & 0.55 & 30 & $10^{8}$ & 1.0 & 500 & 30.78 & 30.80 & 30.78 & 0.147 \\
PRB & 8 & 4 & 2048 & 1024 & 384 & 0.58 & 0.58 & 0.00 & 0.61 & 30 & $10^{8}$ & 1.0 & 600 & 30.75 & 30.75 & 30.82 & 0.123 \\
PRB & 8 & 4 & 2048 & 1024 & 384 & 0.63 & 0.63 & 0.01 & 0.66 & 30 & $10^{8}$ & 1.0 & 700 & 30.72 & 30.73 & 30.69 & 0.107 \\
PRB & 8 & 4 & 2048 & 1024 & 384 & 0.68 & 0.68 & 0.01 & 0.71 & 30 & $10^{8}$ & 1.0 & 800 & 30.58 & 30.59 & 30.57 & 0.0935 \\
PRB & 8 & 4 & 2048 & 1024 & 384 & 0.70 & 0.70 & 0.01 & 0.74 & 30 & $10^{8}$ & 1.0 & 900 & 30.45 & 30.45 & 30.43 & 0.0804 \\
PRB & 8 & 4 & 2048 & 1024 & 384 & 0.75 & 0.75 & 0.01 & 0.78 & 31 & $10^{8}$ & 1.0 & 1000 & 30.34 & 30.35 & 30.29 & 0.0731 \\
PRB & 8 & 4 & 2048 & 1024 & 384 & 0.85 & 0.85 & 0.01 & 0.89 & 31 & $10^{8}$ & 1.0 & 1200 & 30.21 & 30.22 & 30.23 & 0.066 \\
PRB & 8 & 4 & 2048 & 1024 & 384 & 0.96 & 0.96 & 0.01 & 1.00 & 31 & $10^{8}$ & 1.0 & 1500 & 30.01 & 30.01 & 29.95 & 0.0537 \\
PRB & 8 & 4 & 2048 & 1024 & 384 & 1.08 & 1.08 & 0.01 & 1.12 & 31 & $10^{8}$ & 1.0 & 2000 & 29.33 & 29.32 & 29.38 & 0.0379 \\
PRB & 8 & 4 & 2048 & 1024 & 384 & 1.36 & 1.36 & 0.01 & 1.42 & 32 & $10^{8}$ & 1.0 & 3000 & 27.95 & 27.96 & 27.96 & 0.027 \\
PRB & 8 & 4 & 2048 & 1024 & 384 & 1.60 & 1.60 & 0.01 & 1.67 & 33 & $10^{8}$ & 1.0 & 4000 & 26.93 & 26.93 & 26.90 & 0.0211 \\
PRB & 8 & 4 & 2048 & 1024 & 384 & 1.81 & 1.81 & 0.02 & 1.89 & 33 & $10^{8}$ & 1.0 & 5000 & 26.17 & 26.17 & 26.26 & 0.0172 \\
PRB & 8 & 4 & 2048 & 1024 & 384 & 2.01 & 2.01 & 0.02 & 2.10 & 33 & $10^{8}$ & 1.0 & 6000 & 25.62 & 25.59 & 25.60 & 0.0147 \\
PRB & 8 & 4 & 2048 & 1024 & 384 & 2.18 & 2.18 & 0.02 & 2.28 & 34 & $10^{8}$ & 1.0 & 7000 & 25.35 & 25.34 & 25.45 & 0.0128 \\
PRB & 8 & 4 & 2048 & 1024 & 384 & 2.34 & 2.34 & 0.02 & 2.44 & 34 & $10^{8}$ & 1.0 & 8000 & 25.29 & 25.27 & 25.37 & 0.0112 \\
PRB & 8 & 4 & 2048 & 1024 & 384 & 2.51 & 2.51 & 0.02 & 2.62 & 34 & $10^{8}$ & 1.0 & 9000 & 25.30 & 25.27 & 25.34 & 0.0102 \\
PRB & 8 & 4 & 2048 & 1024 & 384 & 2.67 & 2.67 & 0.02 & 2.78 & 33 & $10^{8}$ & 1.0 & 10000 & 25.52 & 25.49 & 25.52 & 0.00931 \\
\hline

\caption[width=\textwidth]{\label{tab:gridcases}Simulations considered in this work. The aspect ratios of the computational domain are given by $\Gamma_x$ in the streamwise direction and $\Gamma_y$ in the spanwise direction. The values of $N_x$, $N_y$, and $N_z$ indicate the number of grid points in the streamwise, spanwise, and wall-normal direction. The grid spacing in wall units in the streamwise and spanwise directions is given by $\Delta x^+$ and $\Delta y^+$ respectively. The wall-normal grid spacing in wall units at the wall and the mid-height is given by $\Delta z_w^+$ and $\Delta z_c^+$, respectively. The number of grid points in the wall-normal direction within the thermal boundary layer is given by $N_{BL}$. $Nu_w$ indicates the value of Nusselt number computed using the average gradient of reduced temperature profile at the walls using equation (\ref{eqn:Nusselt_def}). $Nu_{\epsilon_\theta}$ and $Nu_{\epsilon_u}$ indicate the Nusselt numbers that satisfy the global balance of thermal and kinetic dissipation equations (\ref{eqn:thermal_diss_srb}) and (\ref{eqn:kinetic_diss_srb}).}

\end{longtable}
\end{landscape}

\bibliographystyle{jfm}
\bibliography{literature_turbulence}

\end{document}